# Hall effect anisotropy in the paramagnetic phase of $Ho_{0.8}Lu_{0.2}B_{12}$ induced by dynamic charge stripes

A.L. Khoroshilov[a*], K.M. Krasikov[a], A.N. Azarevich[a,b], A.V. Bogach[a], V.V. Glushkov[a], V.N. Krasnorussky[a,c], V.V. Voronov[a], N.Yu. Shitsevalova[d], V.B. Filipov[d], S. Gabáni[e], K. Flachbart[e], N. E. Sluchanko[a]

[a]*Prokhorov General Physics Institute of the Russian Academy of Sciences, Vavilova 38, 119991 Moscow, Russia*

[b]*Moscow Institute of Physics and Technology (State University), Moscow Region 141700 Russia*

[c]*Vereshchagin Institute for High Pressure Physics of RAS, 14 Kaluzhskoe Shosse, 142190 Troitsk, Russia*

[d]*Institute for Problems of Materials Science, NASU, Krzhizhanovsky str., 3, 03142 Kyiv, Ukraine*

[e]*Institute of Experimental Physics SAS, Watsonova 47, 04001 Košice, Slovakia*

*E-mail: artem.khoroshilov@phystech.edu

A detailed study of charge transport in the paramagnetic phase of $Ho_{0.8}Lu_{0.2}B_{12}$ strongly correlated antiferromagnet was carried out at temperatures 1.9–300 K in magnetic fields up to 80 kOe. Four mono-domain single crystals with different orientation of normal vectors to the lateral surface of $Ho_{0.8}Lu_{0.2}B_{12}$ samples were investigated in order to establish the changes in Hall effect due to the anisotropy, induced by (*i*) the electronic phase separation (dynamic charge stripes) and (*ii*) formation of the disordered cage-glass state below $T^* \sim 60$ K. It was demonstrated that in magnetic fields above 40 kOe directed along the [001] and [110] axes in *fcc* crystals a considerable intrinsic anisotropic positive component $\rho^{an}_{xy}$ appears in addition to the ordinary negative Hall resistivity contribution. The relation $\rho^{an}_{xy} \sim \rho^{an}_{xx}{}^{1.7}$ was found between anomalous components of the resistivity tensor for $\mathbf{H}||[001]$ below $T^* \sim 60$ K, and the power law $\rho^{an}_{xy} \sim \rho^{an}_{xx}{}^{0.83}$ was detected for the orientation $\mathbf{H}||[110]$ at temperatures $T < T_S \sim 15$ K. It is argued that below $T_S \sim 15$ K the anomalous odd $\rho^{an}_{xy}(T)$ and even $\rho^{an}_{xx}(T)$ parts of the resistivity tensor may be interpreted in terms of formation of a large size cluster in the filamentary structure of fluctuating charges (stripes). We assume that these $\rho^{an}_{xy}(\mathbf{H}\|[001])$ and $\rho^{an}_{xy}(\mathbf{H}\|[110])$ components represent the intrinsic (Berry phase contribution) and extrinsic (skew scattering) mechanism, respectively. An additional ferromagnetic contribution to anomalous Hall effect (AHE) for both ordinary and anisotropic components in Hall signal was registered and attributed to the effect of magnetic polarization of $5d$ states (ferromagnetic nano-domains) in the conduction band of $Ho_{0.8}Lu_{0.2}B_{12}$.

1. Introduction.

Numerous fundamental studies of strongly correlated electron systems (SCES) such as manganites [1–4], high-temperature superconducting (HTSC) cuprates [5–8], iron-based superconductors [9–13], chalcogenides [14] etc., allowed to discover a diversity of physical phenomena universal for SCES. Indeed, all these systems are characterized by a complexity of phase diagrams induced by strong phase separation due to structural or electronic instability [15]. The spatial electronic/magnetic inhomogeneity turns out to be directly related to simultaneously active spin, charge, orbital, and lattice degrees of freedom, which are considered by many researchers as factors responsible for the appearance of high-temperature superconductivity in cuprates, as well as for the emergence of colossal magnetoresistance in manganites [5,16–18]. In particular, there are two possible mechanisms of the formation of spatially inhomogeneous ground states in SCES [19]: (1) disorder resulting from phase separation near a first-order metal-insulator transition caused by an external factor [19,20] and (2) frozen disorder in the glass phase with short-range order formed by nanoscale clusters [21–23]. In the second case, one more and among the most significant mechanisms leading to an inhomogeneous glass state in HTSC oxides is the formation of static and dynamic charge stripes [24]. Such structures have been repeatedly observed in HTSC cuprates and nickelates by both direct and indirect methods [25–29].

Studying the effect of spatial charge inhomogeneity on the scattering of charge carriers in SCES is rather difficult due to their complex composition, low symmetry of crystal structure, and high sensitivity to external conditions (pressure, magnetic field, etc., see, e.g., [4]). In this respect, it is convenient to test another model system with strong electronic correlations - rare-earth dodecaborides (RB$_{12}$). Both electronic (dynamic charge stripes) and structural (dynamic Jahn-Teller effect) instabilities take place in RB$_{12}$ compounds with a simple *fcc* lattice (space group $Fm\bar{3}m$, see fig.1a) the stoichiometry of which can be controlled reliably during the crystal growth [30].

Main factors determining the appearance of spatial inhomogeneity which result to symmetry lowering in *fcc* rare earth (RE) dodecaborides are listed below. Firstly, the cooperative dynamic Jahn-Teller effect in the boron sub-lattice leads to lifting of degeneracy of the highest occupied molecular orbitals (HOMO) in B$_{12}$ octahedrons and to the appearance of static structural distortions with related HOMO splitting E$_{JT}$ ~ 500-1500 K [31]. Secondly, reaching the Ioffe-Regel limit near T$_E$ ~ 150 K causes the development of vibrational instability, which leads to an increase in the density of phonon states at T ~ T$_E$ [32]. Thirdly, order-disorder transition to the cage-glass state at T* ~ 60 K causes random displacements of RE ions from central symmetric positions in B$_{24}$ cuboctahedra, which form a rigid covalent boron framework [32]. As a result, the above factors are enhanced by periodic changes of hybridization between the 5*d*(R) and 2*p*(B) states in the conduction band at T < T* due to the dynamic Jahn-Teller effect in RB$_{12}$ and induce high-frequency charge density fluctuations with frequencies ν$_S$ ~ 240 GHz [33] (dynamic charge stripes, see fig.1a) along one of the <110> directions in the *fcc* structure [34–37]. Moreover, a large size cluster is formed in this stripe arrangement at temperatures T$_S$~ $h$ν$_S$/$k_B$~15 K. The resulting structural and electronic instabilities initiate nanoscale phase separation and cause inevitably strong anisotropy of charge transport in an external magnetic field both in the nonmagnetic reference LuB$_{12}$ [32] and magnetic RB$_{12}$ [36–41]. In particular, aforementioned features of the crystal and electronic structure of RB$_{12}$ have a decisive effect on the characteristics of charge transport upon partial replacement of Lu ions (filled *f*-shell, 4f$^{14}$) by magnetic Ho (4f$^{10}$) ions in Ho$_x$Lu$_{1-x}$B$_{12}$ compounds.

The spatial inhomogeneity of fluctuating electron density is the origin for the strong anisotropy of magnetic phase diagrams in these systems (see, for example, figs.1b-1d for Ho$_{0.8}$Lu$_{0.2}$B$_{12}$ and [39–42]). Indeed, strong magnetic anisotropy is observed, for example, in Ho$_x$Lu$_{1-x}$B$_{12}$ with a high concentration of magnetic ions not only in the paramagnetic state (fig.1e), but also the angular antiferromagnetic phase diagrams reveal a Maltese cross symmetry (see figures 1f-1g, [39,40,42] and also [41] for TmB$_{12}$). It is worth noting, that alike in the nonmagnetic reference compound LuB$_{12}$, strong charge transport anisotropy is observed in the paramagnetic state of Ho$_x$Lu$_{1-x}$B$_{12}$, ErB$_{12}$ [40] and TmB$_{12}$ [41] (see, for example, fig.1e and [43]) and attributed to interaction of electron density fluctuations (stripes) with external magnetic field (for recent review see [44] and references therein).

Until now, studies of electron transport in Ho$_x$Lu$_{1-x}$B$_{12}$ compounds have been mainly focused on transverse magnetoresistance measurements (see, e.g., [38,42,43]). Nevertheless, the recent study of LuB$_{12}$ [45] and the initial short research on Ho$_{0.8}$Lu$_{0.2}$B$_{12}$ (Ref. [46]) have demonstrated a significant anisotropy of the Hall effect due to an anomalous positive anisotropic contribution which appeared below T*. Thus, it is of great interest to study in detail the effect of electronic phase separation on the off-diagonal component of the resistivity tensor in model magnetic compound Ho$_{0.8}$Lu$_{0.2}$B$_{12}$ with dynamic charge stripes. Following the short study [46] this work presents results and detailed analysis of the normal and anomalous contributions to the Hall effect in the paramagnetic state of Ho$_{0.8}$Lu$_{0.2}$B$_{12}$. The results are based on angular and magnetic field measurements of Hall resistivity and determine the anisotropic component of the resistivity tensor for this model system with electronic phase separation (dynamic charge stripes). The observed complex angular behavior of the anisotropic Hall resistivity component attributed to interaction of the filamentary structure of fluctuating charges with external magnetic field. The presented arguments favor the intrinsic mechanism of the formation of the anomalous Hall effect.

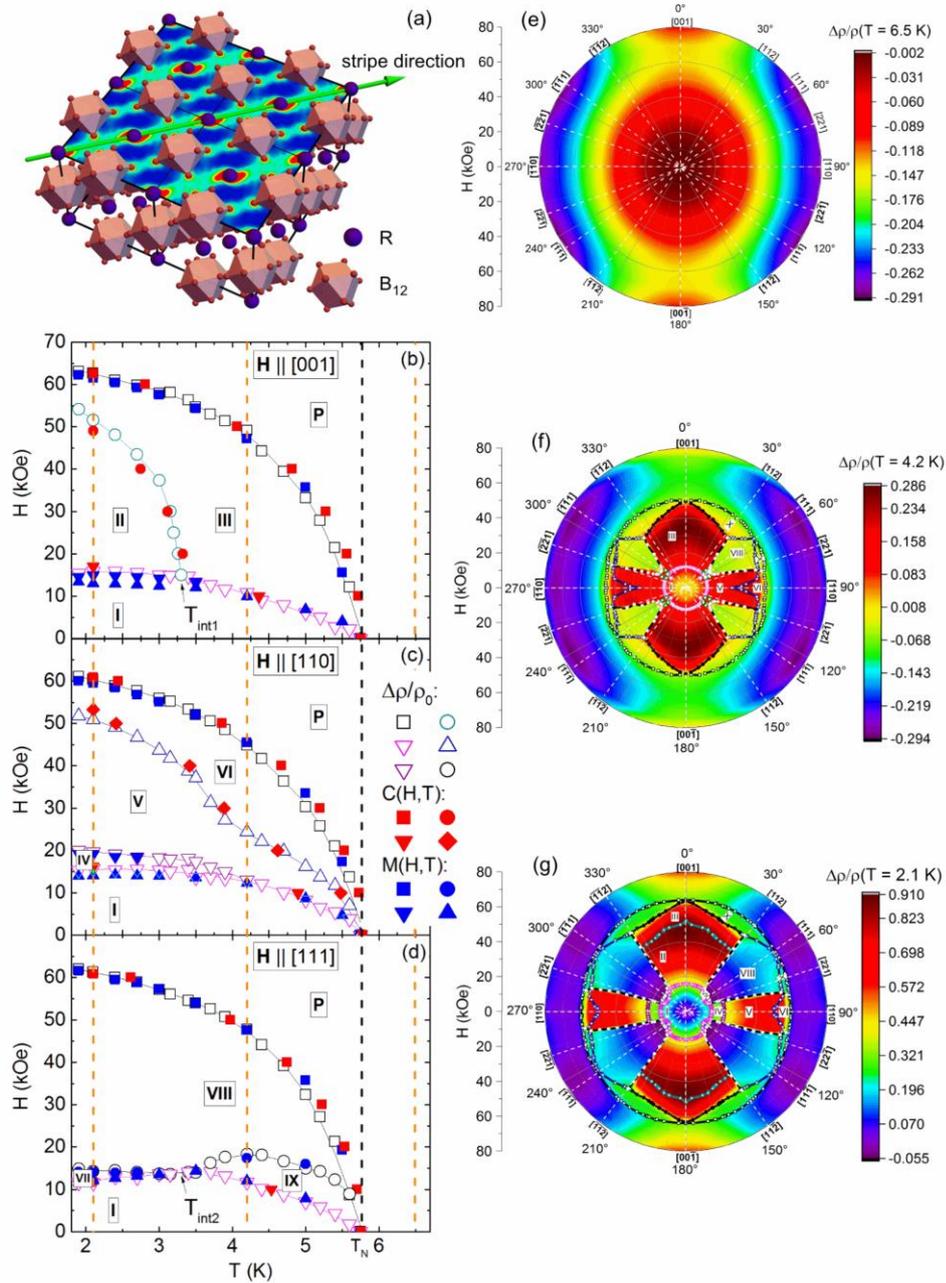

**Fig.1.** (a) Sketch of charge stripes arrangement (green lines) in the $RB_{12}$ crystal structure, (b) - (d) H-T magnetic phase diagrams of the AF state of $Ho_{0.8}Lu_{0.2}B_{12}$ in different field directions, (e) polar plot of the field-angular magnetoresistance in the paramagnetic state and (f) - (g) angular H-φ magnetic phase diagrams (color shows the magnetoresistance amplitude) of the AF state $Ho_{0.8}Lu_{0.2}B_{12}$ at low temperatures (reproduced from [42]).

## 2. Experimental details.

Ho$_{0.8}$Lu$_{0.2}$B$_{12}$ single crystals were grown by crucible-free induction zone melting in an inert argon gas atmosphere (see, e.g., [30]). To measure magnetization, resistivity, magnetoresistance (MR), and Hall effect, we cut the samples (bars) from single-domain crystals with faces aligned along the specified crystal directions with a declination not exceeding 2°.

The magnetization was measured with the help of a SQUID magnetometer MPMS Quantum Design in fields up to 70 kOe in the temperature range 1.9 - 10 K. The external magnetic field was applied along each of the principal crystallographic axis: **H** ∥ [001], **H** ∥ [110] and **H** ∥ [111].

Studies of resistivity, magnetoresistance, and Hall effect were carried out in an installation for galvanomagnetic measurements using the standard five-probe technique with direct current (DC) commutation. The angular dependences of the transverse magnetoresistance and Hall resistivity were obtained using a measuring cell of original design, which enables the rotation of the vector **H** located in the plane perpendicular to fixed current direction **I** ∥ [1$\bar{1}$0] with a minimum step $\Delta\varphi = 0.4°$ (see the schematic view on the inset in fig. 2a). Measurements were carried out in a wide temperature range 1.9 - 300 K in magnetic fields up to 80 kOe, the angle $\varphi = $ **n^H** between the direction **n** normal to the sample surface and external magnetic field **H** varied in the range $\varphi = $ 0 - 360°. The measuring setup was equipped with a stepper motor which enabled automatic control of sample rotation, similar to that one used in [42]. High accuracy of temperature control ($\Delta T \approx$ 0.002 K in the range 1.9 - 7K) and magnetic field stabilization ($\Delta H \approx$ 2 Oe) was ensured, respectively, by Cryotel TC 1.5/300 temperature controller and Cryotel SMPS 100 superconducting magnet power supply in combination with CERNOX 1050 thermometer and n-InSb Hall sensors.

### 3. Experimental results and data analysis.

### 3.1. Temperature dependences of resistivity and Hall resistivity.

Traditional Hall effect studies calculate the value of Hall coefficient as $R_H = \rho_H/H = [(V_H(+H) - V_H(-H))/(2I)]\cdot d/H$, where I is the current through the sample, *d* the sample thickness (sample size along the normal **n**, see inset in fig.2a), and $V_H(+/-H)$ are the voltages measured from Hall probes in two opposite orientations of the external magnetic field $\mathbf{H} \perp \mathbf{I}$. Taking into account the complex field dependence of Hall effect in parent compound $LuB_{12}$ [45], and because the detected anomalous contributions to Hall resistivity have different origin [47], the term "reduced Hall resistivity" for $\rho_H/H$ will be used below in this work instead of the Hall coefficient $R_H$.

Fig. 2 presents the temperature dependences of resistivity $\rho(T)$ in zero magnetic field and for H=80 kOe, as well as the reduced Hall resistivity $\rho_H(T)/H$ in $Ho_{0.8}Lu_{0.2}B_{12}$ calculated for three principal directions $\mathbf{H}\|\mathbf{n}\|[001]$, $\mathbf{H}\|\mathbf{n}\|[110]$ and $\mathbf{H}\|\mathbf{n}\|[111]$. The vertical dashed lines point to the temperature of cage-glass state formation $T^* \sim 60$ K [32] and to temperatures (below 10K), at which Hall resistivity was studied in more detail (see figs. 5, 6, 9, 11-12 below). In zero magnetic field the $\rho(T)$ curves measured for all three $Ho_{0.8}Lu_{0.2}B_{12}$ samples correspond to metallic conductivity with the same RRR value $\rho(300\ K)/\rho(4.2\ K) = 23.7$ (Fig.2a). The data for H=0 kOe and 80 kOe are clearly separated below $T^* \sim 60$ K, indicating a pronounced sign-alternating magnetoresistance. Note that the $\rho(T, H=80\ kOe)$ curves for crystals with $\mathbf{n}\|[110]$ and $\mathbf{n}\|[111]$ match together above $T_S \sim 15$ K differing noticeably at lower temperatures. On the contrary, at T < $T^* \sim 60$ K the $\rho(T,H=80\ kOe)$ dependence for field direction $\mathbf{H}\|\mathbf{n}\|[001]$ lies well above the ones for $\mathbf{H}\|\mathbf{n}\|[110]$ and $\mathbf{H}\|\mathbf{n}\|[111]$, so the MR anisotropy reaches values $\rho(\mathbf{n}\|[001])/\rho(\mathbf{n}\|[111]) \approx 1.8$ ($\approx 80\%$) at T = 2.1 K in 80 kOe.

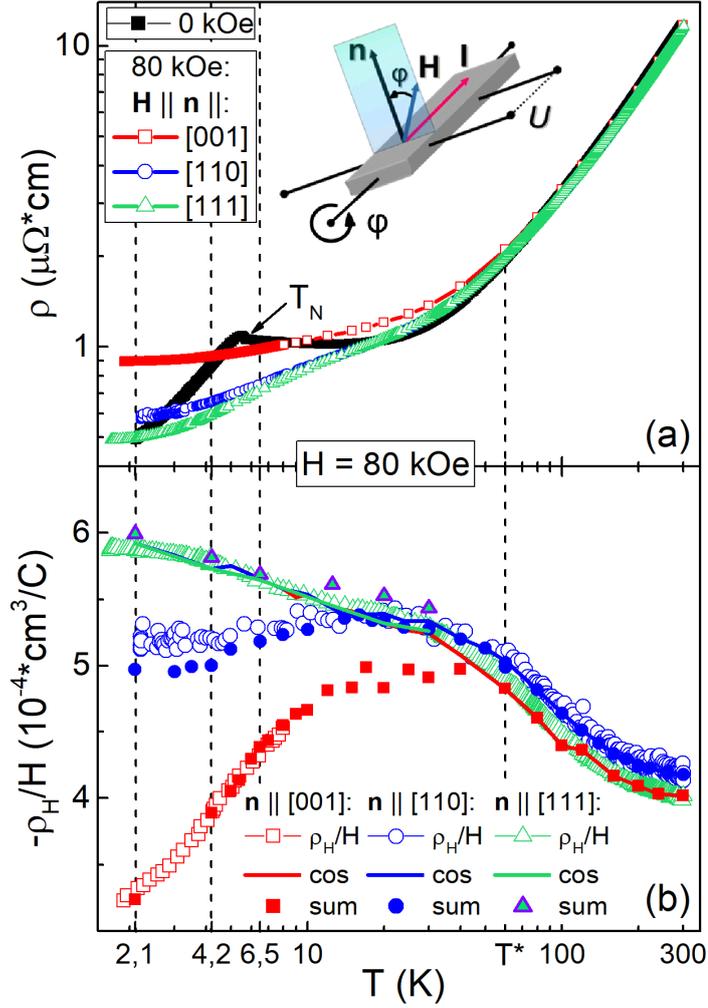

**Fig.2.** Temperature dependencies (a) of resistivity $\rho(T)$ in zero magnetic field and for H=80 kOe, as well as (b) the absolute value of reduced Hall resistivity $-\rho_H(T)/H$ in $Ho_{0.8}Lu_{0.2}B_{12}$ for **H**||**n**||[001], **H**||**n**||[110] and **H**||**n**||[111] (see inset). In panel (b) open symbols and solid lines show the experimental data for $-\rho_H/H$ and the sum of isotropic and anisotropic contributions to Hall effect as *sum* = $\rho_{H0}/H + \rho_H^{an}/H$, correspondingly. Closed symbols show the reduced amplitudes of the isotropic component *cos* = $\rho_{H0}/H$ (see text for detail). Vertical dashed lines point to the temperature of cage-glass state formation T* ~ 60 K [32] and to temperatures (below 10 K) at which Hall resistivity was studied in more detail.

Open symbols in Fig. 2b show the results of $\rho_H/H(T)$ measurements in the scheme with two opposite external field orientations. Significant differences between the $\rho_H/H(T)$ dependencies for different field directions appear below T* ~ 60 K, while the curves for **H**||**n**||[110] and **H**||**n**||[111] start to diverge below 15 K. In this case, the lowest negative values of $\rho_H/H(T)$ are detected for the **n**||[001] sample, while the highest values are observed for **H**||**n**||[111]. In this case,

the maximal anisotropy of the reduced Hall resistivity at T = 2.1 K in field of 80 kOe is comparable with the resistance anisotropy $|\rho_H/H(max)|/|\rho_H/H(min)| \approx 1.83$ ($\approx 83\%$). The temperature dependencies of $\rho_H/H(T)$ allow to identify some anisotropic positive component of the Hall signal which appears in $Ho_{0.8}Lu_{0.2}B_{12}$ in strong magnetic fields. It is worth noting that temperature lowering results in the increase of anisotropy for both Hall resistivity and MR components.

### 3.2. Field dependencies of Hall resistivity and magnetization.

Figs. 3a-3c show the reduced Hall resistivity $\rho_H/B(B)$ vs magnetic induction **B** at temperatures of 2.1, 4.2, and 6.5 K measured in the conventional scheme on three samples with magnetic field **H** applied along **n**∥[001], **n**∥[110] and **n**∥[111]. The related magnetic susceptibility M/B(B) is shown in Figs. 3d-3f. Fig. 3g shows the temperature dependence of magnetic susceptibility M/B(T) ≡ $\chi$(T) in magnetic field H=100 Oe. All the data were corrected by demagnetizing fields. It is seen from Figs. 3d-3f, that M/B(B) decreases with increasing both field and temperature in paramagnetic phase indicating a trend towards saturation of magnetization in strong magnetic fields. It can be discerned from Figs. 3d-3f that in paramagnetic state the M/B(B) curves demonstrate similar dependences, and the magnetic anisotropy M/B(**n**∥[001])/M/B(**n**∥[111])-1 in magnetic field H~70 kOe does not exceed 1.4% even at lowest available temperature 2.1 K. Below we analyze Hall effect by using the same dependence M/B(**n**∥[001]) for all three orientations of applied magnetic field.

The AF-P phase transition at $T_N$ = 5.75 K can be clearly recognized on the temperature dependence of magnetic susceptibility measured at H = 100 Oe (see fig. 4). Above $T_N$ the low field

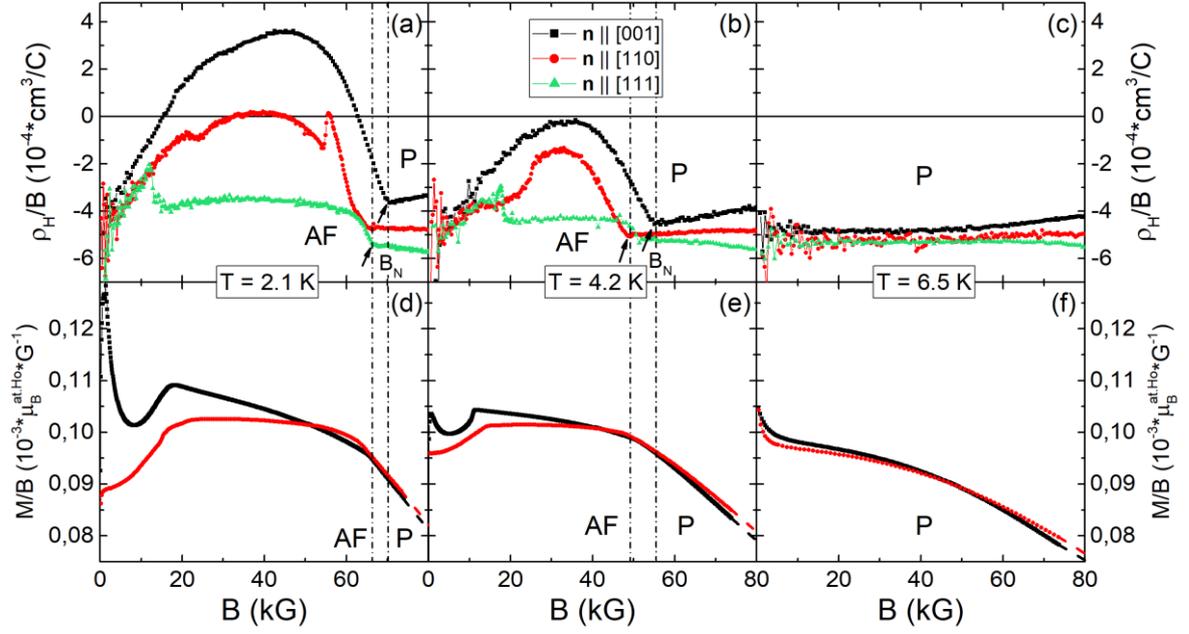

**Fig.3.** (a)-(c) Field dependencies of the reduced Hall resistivity $\rho_H/B$ vs magnetic induction **B** at T= 2.1, 4.2, and 6.5 K for **H**‖**n**‖[001], **n**‖[110] and **n**‖[111], respectively. Arrows at $B_N$ indicate AF-P transitions. **(d)-(f)** Corresponding curves of magnetic susceptibility M/B(B). The dashed lines show the approximation in the interval 7-8 T (see text).

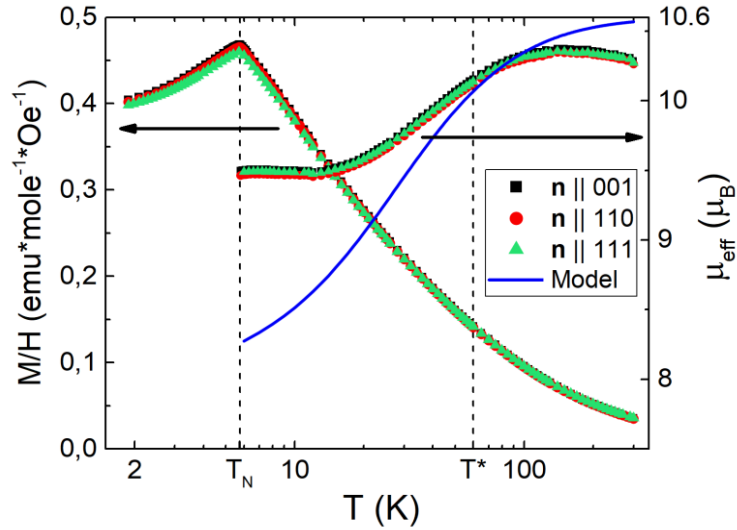

**Fig. 4.** Temperature dependencies of magnetic susceptibility (left scale) and effective moment (right scale) at H = 0.1 kOe for different **H**‖**n** directions: **H**‖**n**‖[001], **n**‖[110] and **n**‖[111] (see text). Solid line indicates the changes of the magnetic moment of Ho $^5I_8$-multiplet splitting by CEF in HoB$_{12}$ (see [48]).

magnetic susceptibility $\chi(T)$ may be described approximately by Curie-Weiss type dependence

$$\chi = M/H = N_{Ho} \mu_{eff}^2 /(3k_B (T-\theta_p)) + \chi_0 \qquad (1),$$

where $N_{Ho} = 0.95 \cdot x(Ho) \cdot 10^{22}$ cm$^{-3}$, and $\mu_{eff} \sim 10$ $\mu_B$ are the concentration of Ho-ions and the effective magnetic moment per Ho-ion, respectively ($\mu_B$ and $k_B$ are Bohr magneton and Boltzmann constant), $\theta_p \approx -14$ K is paramagnetic Curie temperature corresponding to AF exchange between magnetic dipoles and $\chi_0 \approx -1.78 \cdot 10^{-3}$ $\mu_B$/mole/Oe is temperature independent combination of (*i*) diamagnetic susceptibility of the boron cage and (*ii*) Pauli paramagnetism of conduction electrons. Fitting of $\chi(T)$ by Eq. (1) with temperature dependent $\mu_{eff}(T)$ indicates that within the experimental accuracy the susceptibility follows the Curie-Weiss dependence with magnetic moment, which is only slightly below the total moment $\mu_{eff} \approx 10.6$ $\mu_B$ of Ho$^{3+}$ 4*f*-shell in the range 80 - 300 K. As the population of excited magnetic states of the Ho$^{3+}$ $^5I_8$ multiplet splitting by crystalline electric field (CEF) [48] declines significantly in the range 8-80 K, $\mu_{eff}$ decreases moderately (to 9.5 $\mu_B$, see fig.4, right scale), so, even at $T_N$ its value exceeds noticeably the magnetic moment of $\Gamma_5^1$ ground state triplet $\mu_{eff}(\Gamma_5^1) \approx 7.5$ $\mu_B$ (solid line in fig.4, right scale). The difference ($\Delta\mu_{eff} \sim 1.5$ $\mu_B$, fig.4) may be related to ferromagnetic correlations, which develop in this SCES below $T^* \sim 60$ K. Note, that below 25 K various short-range ordering features including ferromagnetic components were previously observed in magnetic RB$_{12}$ [49–51].

As can be seen from figs. 3a-3c, the behavior of reduced Hall resistivity $\rho_H/B(B)$ differs significantly depending on **B** direction. Thus, in the paramagnetic region for **B**||**n**||[001] $\rho_H/B(B)$ turns out to decrease, for **B**||**n**||[110] the curve is practically field independent, and for **B**||**n**||[111], an increase of negative $\rho_H/B(B)$ values is observed. These tendencies persist in temperature range 2.1 - 6.5 K in the paramagnetic phase (above Neel field, B > B$_N$ in figs. 3a-3c), and the anisotropy of $\rho_H/B(\mathbf{n}||[111])/\rho_H/B(\mathbf{n}||[001])$ -1 reaches values of ~ 80% at 2.1 K for B = 80 kG in accordance with the results in fig. 2b. Such large anisotropy is very unusual for the paramagnetic state of *fcc* metals Ho$_{0.8}$Lu$_{0.2}$B$_{12}$ in the presence of strong charge carrier scattering on randomly arranged magnetic Ho$^{3+}$ ions.

### 3.3. Angular dependences of Hall resistivity in the paramagnetic state of $Ho_{0.8}Lu_{0.2}B_{12}$.

To reveal the nature of the strong anisotropy of $\rho_H/H$ Hall resistivity (fig. 2b, fig. 3a-3c) as well as to separate different contributions to Hall effect, it is of interest to study the angular dependencies of Hall resistivity $\rho_H(\varphi)$ in $Ho_{0.8}Lu_{0.2}B_{12}$ for different configurations of external magnetic field with respect to principal crystallographic directions. Here we present precision measurements of Hall resistivity $\rho_H(\varphi,T_0,H_0,\mathbf{n})$ angular dependencies performed at 2.1–300 K in magnetic field up to 80 kOe for four $Ho_{0.8}Lu_{0.2}B_{12}$ crystals with different orientations of the lateral surface: $\mathbf{n}\|[001]$, $\mathbf{n}\|[110]$, $\mathbf{n}\|[111]$ and $\mathbf{n}\|[112]$ (see inset in fig. 2a). Each direction of $\mathbf{n}$ corresponds to a $\mathbf{H}$ orientation in (110) plane which is transverse to DC $\mathbf{I}\|[1\text{-}10]$. For clarity, Fig. S1 in [52] demonstrates a direct correlation between the data obtained in the usual scheme of Hall effect measurements with two opposite directions of $\pm\mathbf{H}\|\mathbf{n}$ and the data extracted with the help of step-by-step rotation of the sample around $\mathbf{I}\|[1\text{-}10]$ with a fixed $\mathbf{H}$ direction in the plane perpendicular to the rotation axis (see the inset in fig. 2a).

Fig. 5 shows the results of angular $\rho_H(\varphi)$ measurements in magnetic field H= 80 kOe for samples with normal directions $\mathbf{n}\|[001]$ and $\mathbf{n}\|[110]$ in the temperature ranges 40 - 300 K (Fig. 5a, 5c) and 2-25 K (Fig. 5b, 5d). The experimental results were fitted by relation

$$\rho_H(\varphi) = \rho_H^{const} + \rho_{H0}\cdot\cos(\varphi+\Delta\varphi) + \rho_H^{an}(\varphi) \qquad (2),$$

where $\rho_H^{const}$ is an angle independent component, $\rho_{H0}$ is the amplitude of the isotropic cosine-like contribution to Hall resistivity $f_{cos}(\varphi) = \rho_{H0}\cdot\cos(\varphi+\Delta\varphi)$, $\Delta\varphi$ is the phase shift, and $\rho_H^{an}(\varphi)=\rho_{H0}^{an}\cdot g(\varphi)$ is the anisotropic contribution to Hall resistivity (see fig. 5). The approximation of $\rho_H(\varphi)$ within the framework of Eq. (2) for two crystals with normal directions

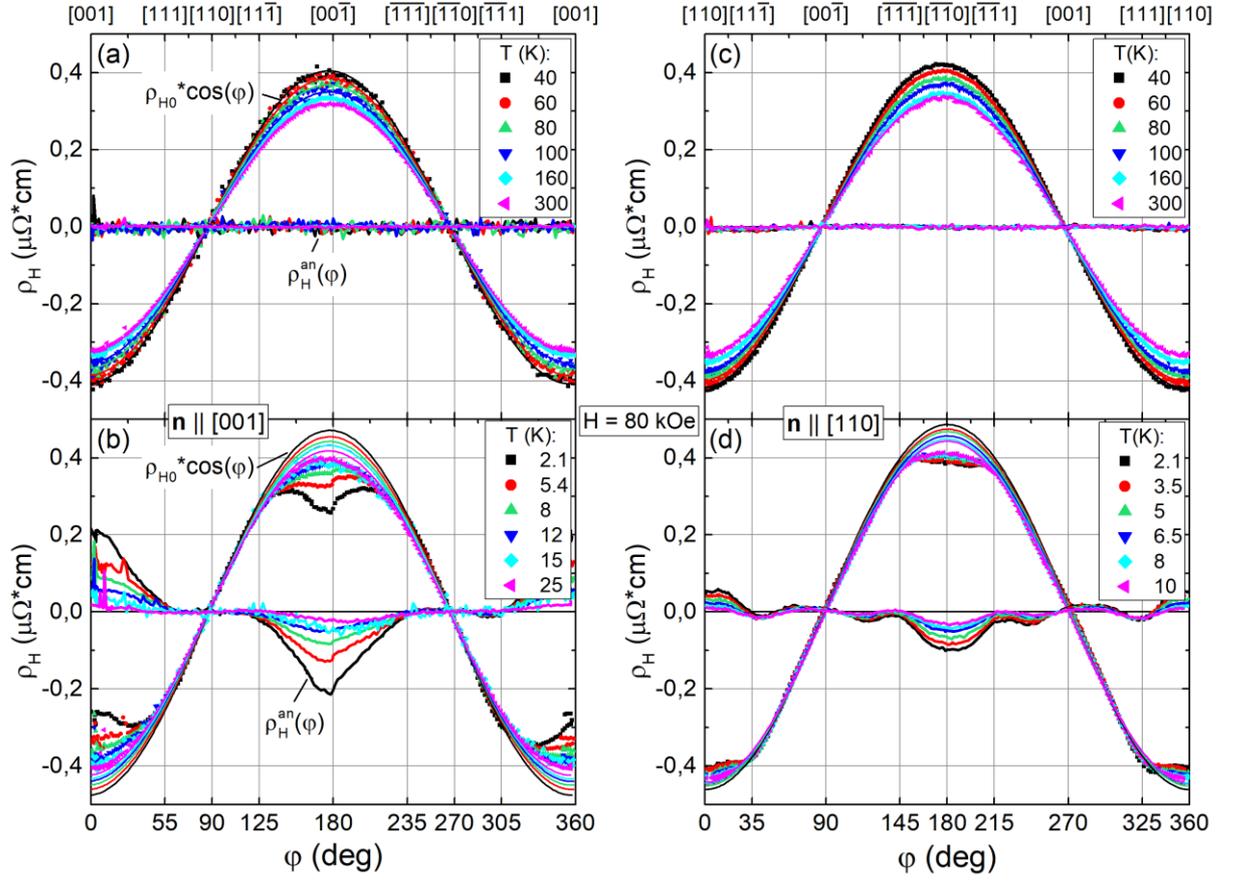

**Fig. 5. (a)-(d)** Angular dependencies of Hall resistivity $\rho_H(\varphi)$ measured in H=80 kOe for the samples with **n**||[001] and **n**||[110] in the temperature range 2-300 K. Symbols show the experimental $\rho_H(\varphi)$ data, thin and thick curves demonstrate the isotropic component $f_{cos}(\varphi) \approx \rho_{H0}\cdot\cos(\varphi)$, and the anisotropic difference $\rho_H^{an}(\varphi) = \rho_H(\varphi) - f_{cos}(\varphi)$, correspondingly.

**n**||[001] and **n**||[110] was carried out in a wide vicinity of zero values of these experimental dependences, in the range of angles $\Delta\varphi=90\pm35°$ and $\Delta\varphi=270\pm35°$. By analogy, $\rho_H(\varphi)$ curves for samples with **n**||[111] and **n**||[112] were approximated in the same intervals $\Delta\varphi=90\pm35°$ and $\Delta\varphi=270\pm35°$ near the zeros of angular dependencies, but without reference to certain crystallographic directions (see Fig. S2 in [52]). As a result, the $\rho_{H0}(T, H, \mathbf{n})$ and $\Delta\varphi(T, H, \mathbf{n})$ parameters of isotropic contribution $f_{cos}(\varphi)$ were found directly from Eq.(2). The anisotropic contribution $\rho_H^{an}(T, H, \mathbf{n})$ was determined as the half-sum of $\rho_H^{an}(\varphi)$ absolute values found along **n** at $\varphi=0°$ and $\varphi=180°$. As can be seen from the analysis of angular $\rho_H^{an}(\varphi)$ dependencies undertaken below, the proposed approach reveals significant limitations and inaccuracies inherent

in measurements of Hall effect by the traditional scheme. Taking into account that $\rho_H^{const}$ and $\Delta\phi \approx$ 3-5° lead only to small corrections in determining $\rho_{H0}=\rho_{H0}(T, H, \mathbf{n})$ and $\rho_{H0}^{an}=\rho_{H0}^{an}(T, H, \mathbf{n})$ amplitudes in Eq. (2), below the experimentally measured Hall resistivity is discussed as a sum of isotropic and anisotropic contributions $\rho_H(\phi) \approx \rho_{H0}\cdot\cos(\phi) + \rho_{H0}^{an}\cos(\phi)\cdot g(\phi)$.

In the range 40–300 K at H=80 kOe the experimental data for $\mathbf{n}\|[001]$ and $\mathbf{n}\|[110]$ samples (Figs. 5a, 5c) are well fitted by a cosine dependence, indicating the absence of anisotropic contribution, and hence $\rho_H^{an}(\phi) \sim 0$. On the contrary below 40 K the $\rho_H^{an}(\phi)$ curves exhibit a broad feature in a wide range of angles around $\mathbf{H}\|\mathbf{n}\|[001]$ limited by <111> axis with a step just at <001> (Fig. 5b), and several peaks of relatively small amplitude may be identified on $\rho_H^{an}(\phi)$ for $\mathbf{n}\|[110]$ (Fig. 5d). The $\rho_H^{an}(\phi)$ curves for $\mathbf{n}\|[111]$ and $\mathbf{n}\|[112]$ in a field of 80 kOe at temperatures 2.1 - 30 K are presented in [52] (see Fig. S2). Note that the $\rho_H(\phi)$ dependencies for $\mathbf{n}\|[111]$ and $\mathbf{n}\|[112]$ being similar to each other, differ from curves recorded for $\mathbf{n}\|[001]$ and $\mathbf{n}\|[110]$, deviating significantly from cosine dependence in a wide range of angles. The anisotropic contribution $\rho_H^{an}(\phi)$ extracted for $\mathbf{n}\|[111]$ and $\mathbf{n}\|[112]$ samples is close to zero near the normal directions and is characterized by two extrema in the interval between the zeros of $\rho_{H0}\cdot\cos(\phi)$ curve (see Fig. S2 in [52]).

Fig. 6 shows an approximation by Eq. (2) of the measured Hall resistivity $\rho_H(\phi)$ at T = 6.5 K in fields up to 80 kOe for four crystals with $\mathbf{n}\|[001]$, $\mathbf{n}\|[110]$, $\mathbf{n}\|[111]$ and $\mathbf{n}\|[112]$. It is seen that the anisotropic contribution $\rho_H^{an}(\phi)$ appears above 40 kOe along the normal direction, has the largest amplitude for $\mathbf{H}\|\mathbf{n}\|[001]$, decreases by a factor of 2 for $\mathbf{H}\|\mathbf{n}\|[110]$ and goes to zero for $\mathbf{H}\|\mathbf{n}\|[111]$ and $\mathbf{H}\|\mathbf{n}\|[112]$ (Fig. 6). We emphasize especially that below 40 kOe, the experimental data (symbols) and cosine fits (thin solid lines) coincide with a good accuracy indicating the absence of an anisotropic component $\rho_H^{an}(\phi)$ in the low field region (see also Fig. S3 in [52] for the $\mathbf{n}\|[111]$ at T = 20 K).

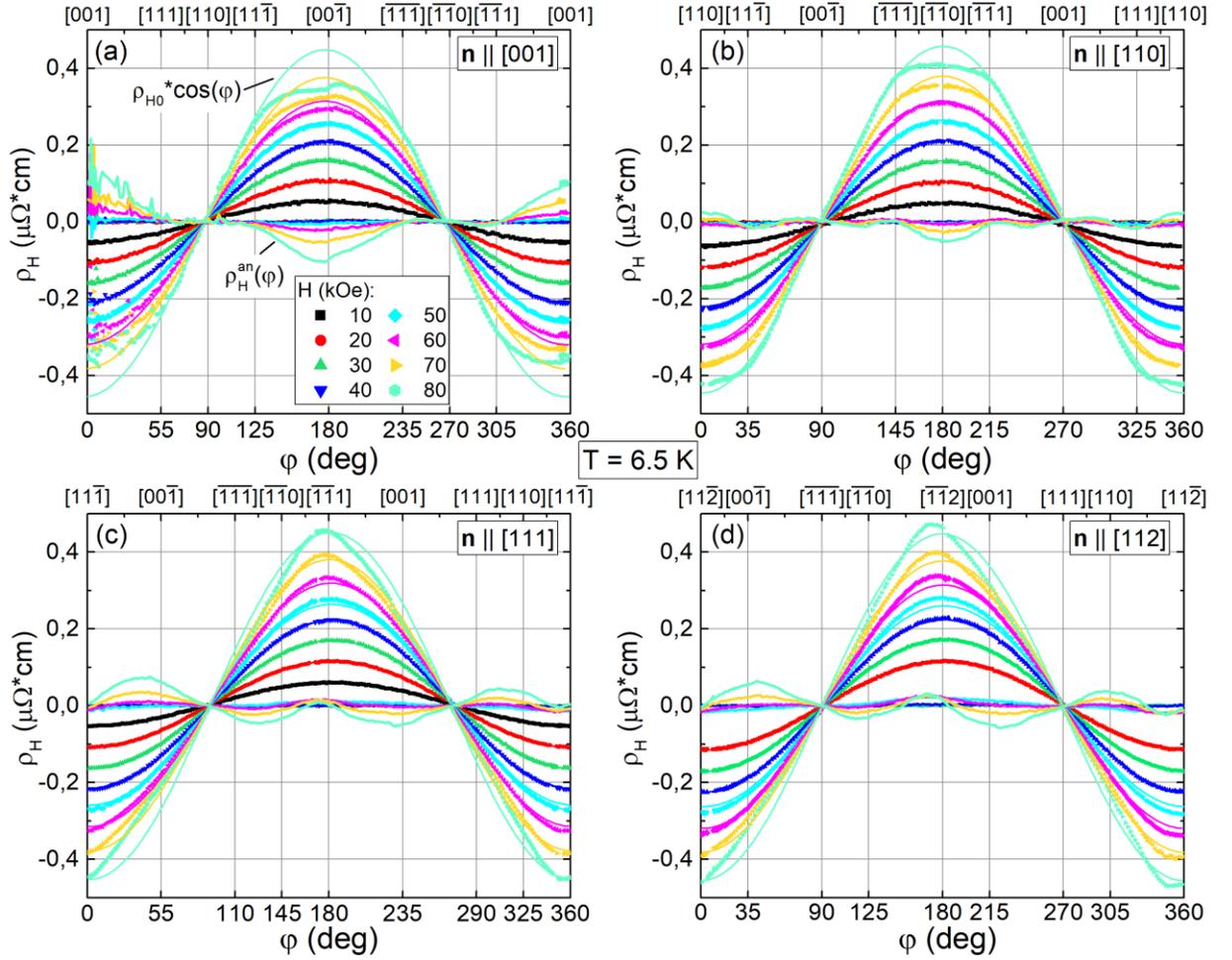

**Fig. 6. (a)-(d)** Angular dependencies of Hall resistivity $\rho_H(\varphi)$ at T= 6.5 K in magnetic field up to 80 kOe for **n**||[001], **n**||[110], **n**||[111] and **n**||[112], correspondingly. Symbols show the experimental $\rho_H(\varphi)$ curves, thin and thick lines indicate the isotropic $f_{cos}(\varphi) \approx \rho_{H0} \cdot \cos(\varphi)$ and anisotropic $\rho_H^{an}(\varphi) = \rho_H(\varphi) - f_{cos}(\varphi)$ contributions, correspondingly.

It is worth noting that in the range T>T*~60 K the temperature dependencies of reduced Hall resistivity $\rho_H/H(T)$ at H=80 kOe for **n**||[001], **n**||[110] and **n**||[111] coincide within our experimental accuracy (Fig. 2b), and the angular $\rho_H(\varphi)$ curves fitted well by cosine (Figs. 5a, 5c), are also close to each other. For **n**||[110] and **n**||[111] samples the $\rho_H(\varphi)$ data coincide in the temperature range of 15–60 K (Fig. 5a and Fig. S2a in [52]) as well and, as a result, the amplitude of $\rho_{H0}$ takes similar values. Below $T_S \sim$ 15 K at H>40 kOe, significant deviations of the $\rho_H/H(T)$ curves from cosines are observed for the crystals with **n**||[110], **n**||[111] and **n**||[112] in intervals $\Delta\varphi=90\pm35°$ and $\Delta\varphi=270\pm35°$ (see, e.g., Fig. 6). In our opinion, this observation allows to explain

the different trends of $\rho_H(T)$ data (Fig. 2) proving a new approach for eliminating large errors in the $\rho_{H0}$ amplitude of the main component as estimated in the commonly used $\pm\mathbf{H}$ scheme. Indeed, at low temperatures and in magnetic fields H>40 kOe the assumption of isotropic normal contribution to Hall effect allows us to use a single, common $\rho_{H0}$ value, found from the analysis in Eq. (2) for $\mathbf{n}\|[001]$, for all four studied crystals with different $\mathbf{n}$ directions. At the same time, in low magnetic fields H ≤ 40 kOe and for temperatures below $T_S \sim$ 15 K, the $\rho_H(\varphi)$ curves for $\mathbf{n}\|[001]$ differ only slightly from cosine in the intervals $\Delta\varphi=90\pm35°$ and $\Delta\varphi=270\pm35°$ (see Fig. 6). Therefore, approximation by Eq. (2) was carried out with the isotropic contribution using the $\rho_{H0}$ values detected for each of these four samples.

### 3.4. Analysis of contributions to Hall resistivity.

The fitting parameters of $\rho_H(\varphi)$ dependencies obtained from the analysis (see Eq. (2), figs. 5-6 and [52]) in the paramagnetic phase of $Ho_{0.8}Lu_{0.2}B_{12}$ are shown in fig. 7. Different symbols correspond to isotropic $\rho_{H0}/H(T_0,H)$ and anisotropic $\rho_H^{an}/H(T_0,H)$ components estimated at $T_0$ = 2.1, 4.2, and 6.5 K. The data for different samples with $\mathbf{n}\|[001]$, $\mathbf{n}\|[110]$, $\mathbf{n}\|[111]$ and $\mathbf{n}\|[112]$ in fig. 7 are indicated by different colors. It is seen that at T = 6.5 K for $\mathbf{n}\|[001]$ the value of $\rho_{H0}/H$ is practically constant below 40 kOe (fig. 7a), while in the range H > 40 kOe, the absolute values of $\rho_{H0}/H(T_0,H)$ increase linearly. When temperature decreases, the high field linear dependence shifts down to the region of larger negative values. For $\mathbf{n}\|[110]$, $\mathbf{n}\|[111]$ and $\mathbf{n}\|[112]$ samples, the absolute values of $\rho_{H0}/H(T_0,H)$ decrease moderately with increasing of magnetic field below 40 kOe. It is worth noting that in the paramagnetic state the variation of the isotropic $\rho_{H0}/H(T_0,H)$ component may be attributed to significant (~14%) and non-monotonous change of the concentration of conduction electrons if we assume one type of charge carriers (see right axis in fig. 7a). Note also that the amplitude of anisotropic contribution $\rho_H^{an}/H(T_0,H)$ turns out to be small and negative below 40 kOe. In stronger magnetic field H > 40 kOe this anisotropic component increases for samples with $\mathbf{n}\|[001]$ and $\mathbf{n}\|[110]$ (fig. 7b), with the amplitude $\rho_H^{an}/H(H)$ for $\mathbf{n}\|[001]$ being more than 2 times higher than the corresponding values for **n**||[110], while for **n**||[111] and **n**||[112], small negative values turn out to be close to zero (see also Figs. S2 and S5 in [52]). Thus, the behavior of the second, anisotropic $\rho_H^{an}/H$ component of the reduced Hall resistivity also shifts qualitatively, indicating a dramatic changes in charge transport regime (Fig. 7).

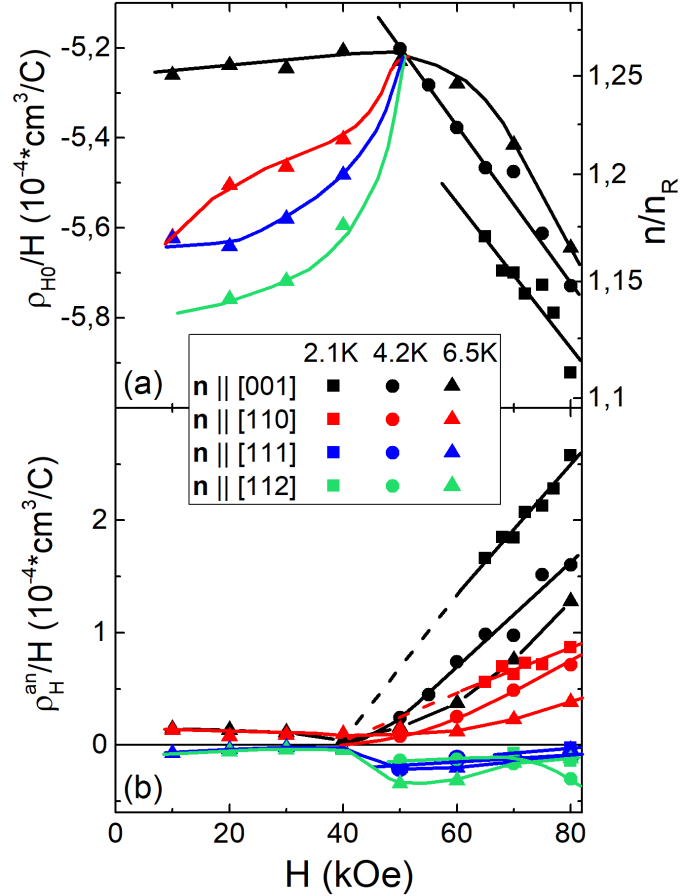

**Fig. 7.** Reduced amplitudes of **(a)** isotropic $\rho_{H0}/H$ and **(b)** anisotropic $\rho_H^{an}/H$ contributions *vs* external magnetic field H at temperatures of 2.1, 4.2, and 6.5 K for samples with **n**||[001], **n**||[110], **n**||[111] and **n**||[112]. Symbols of different shapes and colors indicate different temperatures and samples with different **n** directions.

The experimental $\rho_H/H(T,H_0=80$ kOe) data obtained in the traditional, commonly used technique with two opposite orientations of applied magnetic field ±**H**||**n**, from one side, and the isotropic component found in angular dependencies of Hall resistivity $\rho_{H0}/H(T,H_0)$, from the other, are compared in Fig. 2b. It can be seen, that the $\rho_{H0}/H(T,H_0)$ data coincide with good accuracy with the $\rho_H/H$ values detected in traditional measurements for **n**||[111] at all investigated

temperatures 1.9 - 300 K. For the sample with **n**∥[110] $\rho_{H0}/H$ starts to deviate from $\rho_{H0}/H(T,H_0)$ at $T<T_S$ ~15 K, and for **n**∥[001] noticeable differences arise already upon the transition to the cage-glass state below $T^* \sim 60$ K (Fig. 2b). This observation allows to attribute the appearance of the strong anisotropy of the Hall effect in $Ho_{0.8}Lu_{0.2}B_{12}$ to the additional contribution $\rho_H^{an}/H$. Fig. 2b shows also a comparison of the parameter *sum* = $\rho_{H0}/H + \rho_H^{an}/H$, which corresponds to Hall effect amplitude detected from $\rho_H/H$ angular dependencies, with the reduced Hall resistivity measured in traditional experiment from two field dependencies for ±**H**∥**n**. It can be seen that the temperature behavior of the *sum* is in good agreement with the experimental $\rho_H/H$ dependencies for all the crystals under investigation (**n**∥[001], **n**∥[110] and **n**∥[111]).

It can be discerned from Fig. 2b and Fig. 7, that the isotropic contribution $\rho_{H0}(\varphi)$ is dominant in fields below 40 kOe for all samples under investigation. For H = 80 kOe Hall effect becomes isotropic above $T^* \sim 60$ K for the sample with **n**∥[001] and above $T_S \sim 15$ K for **n**∥[110], while for **n**∥[111] the isotropic contribution $\rho_{H0}(\varphi)$ dominates in the entire temperature range. As a result, using the relation $\rho_{H0}/H(T)= R_H(T) \sim 1/n_e e$ ($e$ is the electron charge) within one type charge carriers, one can estimate the concentration of electrons in the conduction band $n_e$. Fig. 8a shows the Arrhenius plot for the dependencies of the reduced Hall concentration $n_e/n_R$ for **n**∥[001] and **n**∥[110] samples estimated at H= 80 kOe ($n_R = 0.95*10^{22}$ cm$^{-3}$ is the concentration of rare-earth Ho and Lu atoms) and Fig.7a (right axis) demonstrates the field dependence of $n_e/n_R$.

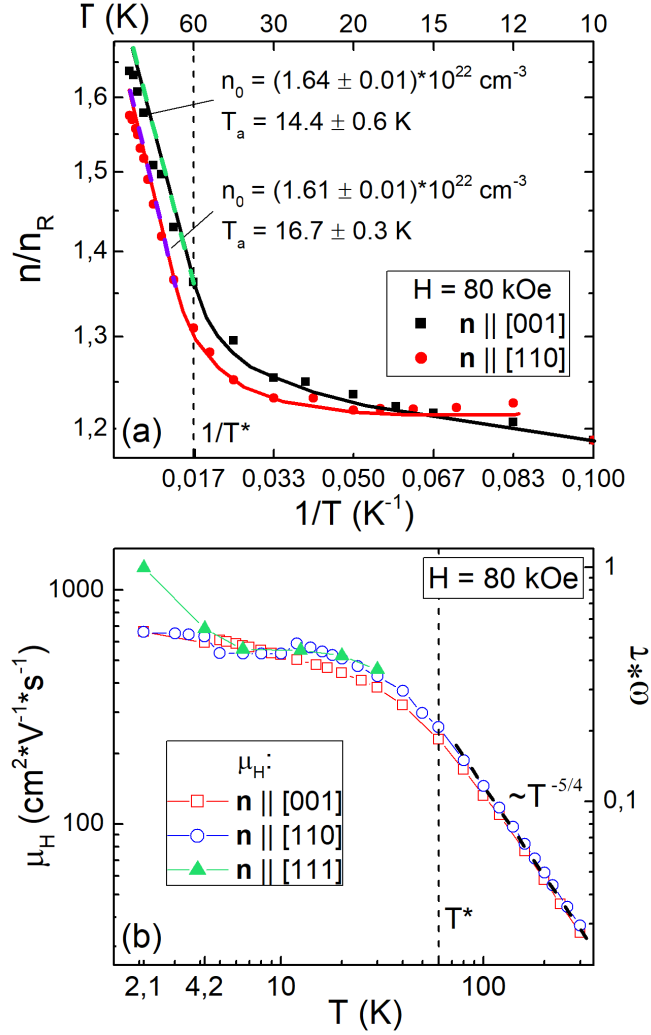

**Fig. 8.** (a) Arrhenius plot $lg(n/n_R) = f(1/T)$ of the reduced Hall concentration for **n**||[001] and **n**||[110] at H=80 kOe. (b) Temperature dependencies of the Hall mobility $\mu_H$ and the parameter $\omega_c\tau \approx \mu_H \cdot H$ at H= 80 kOe for three directions **n**||[001], **n**||[110] and **n**| [111]. The dashed lines show the (a) activation behavior and (b) power law.

It can be seen that the Arrhenius plot provides us by linear $lg(n/n_R) = f(1/T)$ dependencies above the cage-glass transition temperature T* ~ 60 K. Within the framework of relation $1/R_H(T) \sim e \cdot n_0 \cdot \exp(-T_a/T)$, we can estimate the activation temperature $T_a$ and the concentration of charge carriers $n_0$. The extracted values of $T_a$= 14.4 ± 0.6 K for **n**||[001] and $T_a$= 16.7 ± 0.3 K for **n**||[110] are close to $T_S$ ~15 K, and the concentration $n_0 = (1.61-1.64) \cdot 10^{22}$ cm$^{-3}$ coincides within the experimental accuracy for these two crystals. As temperature increases, the reduced Hall concentration changes in the range $n_e/n_R(T) \sim 1.2 - 1.6$ (Fig. 8a) in accordance with the results of Hall effect measurements in rare earth dodecaborides $RB_{12}$ (R = Lu, Tm, Ho, Er) [53].

Fig. 8b shows the temperature dependences of Hall mobility $\mu_H(T) \approx \rho_{H0}(T)/[H \cdot \rho(T)]$ (left scale) and the related parameter $\omega_c\tau \approx \mu_H \cdot H$ (right scale, $\omega_c$ is the cyclotron frequency, $\tau$ is the carrier relaxation time) in magnetic field H = 80 kOe for samples with **n**||[001], **n**||[110] and **n**||[111]. At low temperatures the obtained $\mu_H(T)$ data tend to constant values $\mu_H \sim$ 600-700 cm$^2$/(V·s) (Fig. 8b), in the range T > T* ~ 60 K Hall mobility follows the power law $\mu_H \sim T^{-\alpha}$ with the single exponent being estimated as $\alpha \approx 5/4$ for **n**||[001] and **n**||[110]. Similar behavior of Hall mobility was observed in the range 80–300 K for various $LuB_{12}$ crystals; the $\alpha \approx 7/4$ exponent was detected for samples with large values of RRR≡$\rho$(300 K)/$\rho$(6 K) = 40–70 and $\alpha \approx 3/2$ was estimated for $LuB_{12}$ with a small enough RRR∼12 [45]. The $\alpha = 3/2$ exponent is typical for the scattering of conduction electrons by acoustic phonons (deformation potential) and the increase of $\alpha$ values up to 7/4 in best $LuB_{12}$ crystals was interpreted [45] in terms of charge carriers scattering on both the quasilocal vibrations of RE ions and the boron optical phonons [54] in the presence of JT distortions and the rattling modes of RE ions [51,55,56]. In the case of $Ho_{0.8}Lu_{0.2}B_{12}$ with RRR~10 the decrease of $\alpha$ value from 3/2 to 5/4 could be attributed to emergence of strong magnetic scattering in this dodecaboride with $Ho^{3+}$ magnetic ions. Note that the inequality $\omega_c\tau < 1$ (Fig. 8b), which is fulfilled in the entire temperature range 1.9 - 300 K in fields up to 80 kOe, corresponds to the low field limit for $Ho_{0.8}Lu_{0.2}B_{12}$, indicating that the results of Hall effect measurements are insensitive to the Fermi surface topology.

It is obvious, that the Hall effect in $Ho_{0.8}Lu_{0.2}B_{12}$ is strongly modified by the positive anisotropic contribution $\rho_H^{an}/H$. Figs. 9a, 9b show the angular dependencies of $\rho_H^{an}(\varphi)/H$ at H= 80 kOe for samples with **n**||[001] and **n**||[110]. The same contributions for **n**||[111] and **n**||[112] are shown in Fig. 10. As can be seen from Fig. 9-10, that the $\rho_H^{an}(\varphi)/H$ curves differ in both the amplitude and shape of angular dependence. Since very small changes of $\rho_H^{an}(\varphi)/H$ with temperature are detected for samples with **H**||**n**||[111] and **H**||**n**||[112] in the field along normal

directions (φ = 180° in Fig. 10), the temperature dependence of the anomalous component is only analyzed only for **n**||[001] and **n**||[110].

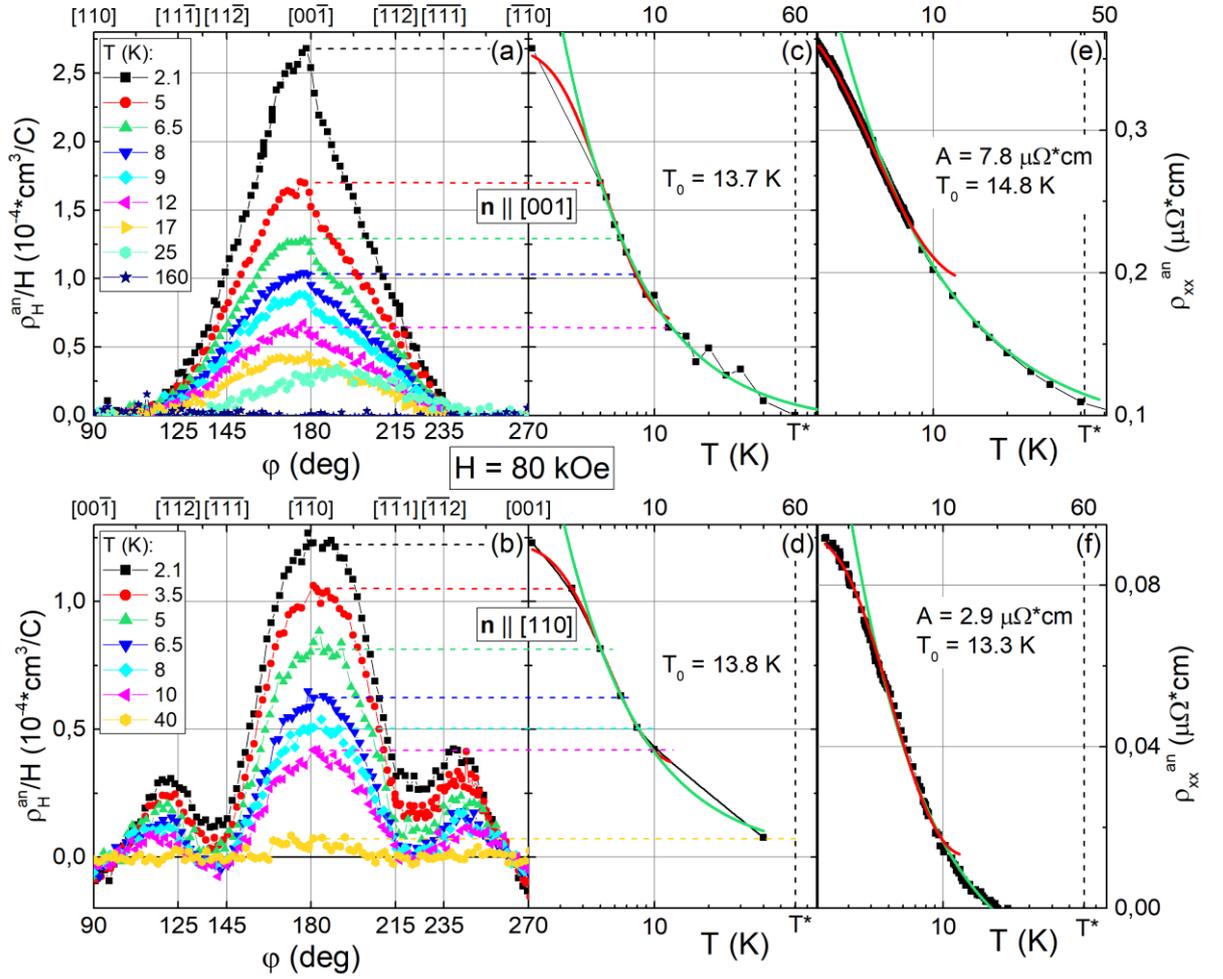

**Fig. 9. (a)-(b)** Anisotropic positive contribution $\rho_H^{an}(\varphi)/H$ for the samples with **n**||[001] and **n**||[110]. **(c)-(d)** Temperature dependencies of the amplitudes $\rho_H^{an}/H$ for **n**||[001] and **n**||[110] in the logarithmic scale. Panels **(e)-(f)** show the temperature dependencies of the anisotropic contribution to resistivity $\rho_{xx}^{an}=\rho(\mathbf{n}||[001], T_0, H=80 \text{ kOe})-\rho(\mathbf{n}||[111], T_0, H=80 \text{ kOe})$ and $\rho_{xx}^{an}= \rho(\mathbf{n}||[110], T_0, H=80 \text{ kOe}) - \rho(\mathbf{n}||[111], T_0, H=80 \text{ kOe})$, respectively. Green and red solid lines on panels **(c)-(f)** show the approximation by Eqs. (3), and (4) (see text). All results correspond to measurements at H= 80 kOe.

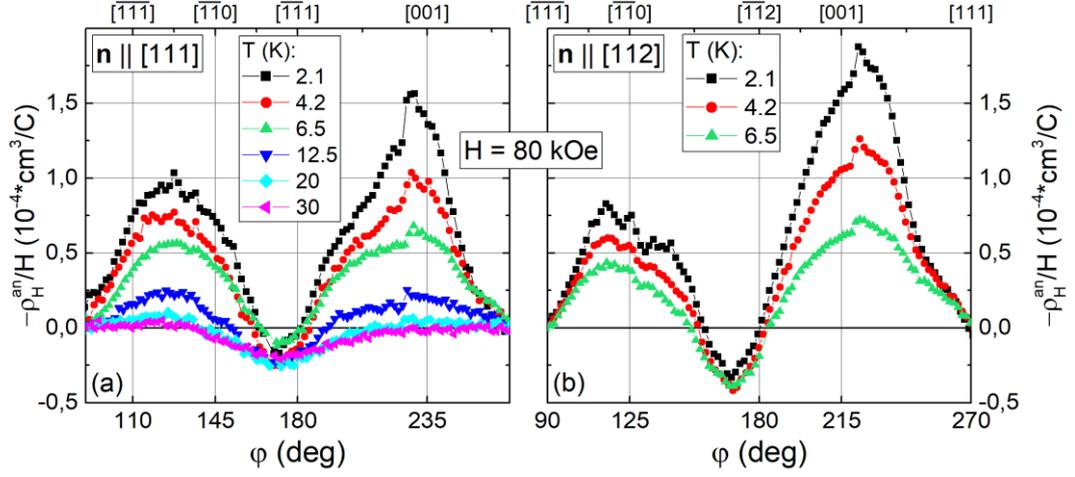

**Fig. 10. (a)-(b)** Anisotropic contributions $\rho_H^{an}(\varphi)/H$ in magnetic field H=80 kOe for the samples with **n**||[111] and **n**||[112], respectively.

The variation of $\rho_H^{an}/H$ with temperature is shown in Figs. 9c, 9d in the logarithmic plot. It is worth noting that two phenomenological relations

$\rho_H^{an}/H \approx C^* \cdot (1/T - 1/T_E)$ (3)

$\rho_H^{an}/H \approx (\rho_H^{an}/H)_0 - A_H \cdot T^{-1} \cdot \exp(-T_0/T)$ (4a)

were used [45,57,58] for the Hall effect analysis in strongly correlated electronic systems with a filamentary structure of conducting channels. Among these, a hyperbolic dependence (3) of Hall resistivity was observed previously in SCES as $CeCu_{6-x}Au_x$ [58] and $Tm_{1-x}Yb_xB_{12}$ [57]. It was considered by authors as a consequence of the appearance of a transverse even component of the Hall signal associated with the formation of stripes [33,44] on the surface and in the bulk of the crystal, similar to nanostripes formed by chains in the normal phase of HTSC [59]. Eq. (4a) was applied to discuss the temperature induced destruction of the coherent state of stripes in $LuB_{12}$ [45]. Here we use equations (3) and (4a) to emphasize the quantitative difference between two various orientations of the magnetic field relative to the crystallographic.

Indeed, the approximation of data in Fig. 9c by Eq. (3) in the temperature range 5 - 40 K results into the values $T_E \approx 132$ K and $C^* \approx 8.9 \cdot 10^{-4}$ cm$^3$/C for the sample with **n**||[001], while for **n**||[110], the value of the parameter $C^* \approx 4.1 \cdot 10^{-4}$ cm$^3$/C turns out to be about half the size (green

curves in Figs. 9c, 9d). Moreover, our analysis of the data in Figs. 9c, 9d (Eq. (4a), red curves) provide $T_a \approx 13.5\pm0.3$ K for these field directions, in agreement with the characteristic temperature of stripe formation $T_S \sim 15$ K in rare-earth magnetic dodecaborides [44], from one side, and with the activation energy $E_a/k_B = T_a \sim 14 - 16$ K found for the main contribution to Hall effect in the interval $T \geq T^*$ (see Fig. 7a), from the other one. It can be seen from Figs. 9c, 9d that for the case $\mathbf{n}\|[001]$ and $\mathbf{n}\|[110]$, Eq. (4a) provides a good description of the experimental curves $\rho_H^{an}/H(T)$ curves at temperatures up to 10 K. However, above 10 K, these fits (see red curves in Figs. 9c, 9d) differ sharply from experiment, indicating the restriction of the phenomenological approach applied. The $A_H$ coefficients in Eq. (4a) differ for these two normal directions by more than 2 times ($A_H = 73.8\cdot10^{-4}$ cm$^3$/C for $\mathbf{n}\|[001]$ and $A_H = 32.3\cdot10^{-4}$ cm$^3$/C for $\mathbf{n}\|[110]$) that is in good agreement with the amplitude ratio for $\rho_H^{an}/H$ (Figs 9c, 9d). Further, similar to the approach developed in [38,45] for LuB$_{12}$, an analysis of the anisotropic positive contribution to magnetoresistance in Figs. 9e-9f for Ho$_{0.8}$Lu$_{0.2}$B$_{12}$ is carried out within the following relation

$$\rho_{xx}^{an}(\mathbf{n},T_0,H=80\text{ kOe}) \approx (\rho_{xx}^{an})_0 - A_{xx}\cdot T^{-1}\cdot\exp(-T_a^\rho/T) \qquad (4b)$$

for two orientations of the external field $\mathbf{H}\|[001]$ and $\mathbf{H}\|[110]$. For each of the samples with $\mathbf{H}\|[001]$ and $\mathbf{H}\|[110]$ the anisotropic component $\rho_{xx}^{an}(\mathbf{n},T_0,H=80\text{ kOe})$ was determined by subtracting from the experimental resistivity data (e.g, $\rho(\mathbf{n}\|[001],T_0,H=80\text{ kOe})$ for $\mathbf{H}\|\mathbf{n}\|[001]$) the dependence $\rho(\mathbf{n}\|[111],T_0,H=80\text{ kOe})$ for $\mathbf{H}\|\mathbf{n}\|[111]$, where the magnetoresistance is minimal. The temperatures $T_a^\rho = 13.3 - 14.8$ K found from this approximation in the same range $T \leq 10$ K turn out to be close to $T_a \approx 13.3 - 13.8$ K and also to $T_S \sim 15$ K (characteristic of the stripe system in rare-earth magnetic dodecaborides [33,44], as well as with the activation temperature $T_a \sim 14 - 16$ K obtained from the Arrhenius-type analysis of the Hall coefficient Ho$_{0.8}$Lu$_{0.2}$B$_{12}$ in the range $T > T^* \sim 60$ K (Fig. 8a).

**4. Discussion.**

## 4.1. Multicomponent analysis of contributions to anomalous Hal effect (AHE) in the regime of ferromagnetic fluctuations.

Previous measurements of the Hall effect in $Ho_{0.5}Lu_{0.5}B_{12}$ in the paramagnetic phase ($T>T_N \approx 3.5$ K) were carried out in traditional scheme with two opposite directions of magnetic field $\pm\mathbf{H}\|\mathbf{n}$ [60]. Normal and anomalous components of the Hall effect observed in [60] were described by the general relation

$$\rho_H/B = R_{H0} + R_S \cdot 4\pi M/B \qquad (5),$$

which is applied usually to AHE in ferromagnetic metals [47,61] ($R_{H0}$ and $R_S$= const(T) are the normal and anomalous Hall coefficients, respectively). According to [47], the ferromagnetic AHE mode represented by Eq. (5) corresponds to the intrinsic scattering mechanism. Taking into account that short-range order effects are observed in the paramagnetic phase of magnetic $RB_{12}$ in the temperature range at least up to $3T_N$ [49,51,62] (see also Fig. 4), and ferromagnetic component was detected in the magnetically ordered state of $HoB_{12}$ in magnetic fields above 20 kOe, it is of interest to perform the analysis within the framework of Eq. (5) of the normal and AHE components in the immediate vicinity of $T_N$. In this case, relying on the above results of angular measurements of $Ho_{0.8}Lu_{0.2}B_{12}$ (Fig. 5-7, 9-10), one should use isotropic $\rho_{H0}(T,H,n)$ and anisotropic $\rho_H^{an}(\varphi,T,H,n)$ components of the Hall signal separated within the framework of Eq. (2) (see Fig. 2-10). It is worth noting that the analysis performed in [60] within the framework of Eq. (5) is applied to the field dependences of Hall resistivity $\rho_H(H)$ obtained in the traditional scheme $\pm\mathbf{H}\|\mathbf{n}$, leading obviously to mixing of contributions $\rho_{H0}(H)$ and $\rho_H^{an}(H)$. As a result, the coefficients $R_{H0}$ and $R_S$ have been determined in [60] for the total averaged Hall resistivity which contains both the isotropic $\rho_{H0}(T, H, n)$ and anisotropic $\rho_H^{an}(\varphi, T, H, n)$ components. Actually, each of the angular contributions $\rho_{H0}$ and $\rho_H^{an}(\varphi)$ is characterized by two independent coefficients $R_0$ and $R_S$, which generally differ in sign. Below we develop the analysis, considering two ferromagnetic components included in the AHE. To keep the generality, we analyze below our Hall effect data for all three principal directions of external magnetic field ($\mathbf{H}\|[001]$, $\mathbf{H}\|[110]$ and **H**∥[111]), despite the fact that for **H**∥[111] the intrinsic AHE is found to be practically negligible (see Fig. 10).

Thus, for a full description of the Hall effect in Ho$_{0.8}$Lu$_{0.2}$B$_{12}$ we used the following relations

$$\begin{cases} \rho_H^0 / B = R_H^0 + R_M^0 \cdot 4\pi M / B \\ \rho_H^{an} / B = R_H^{an} + R_M^{an} \cdot 4\pi M / B \end{cases} \quad (6),$$

where $\rho_H^0/B = \rho_{H0}/B(H, T_0, n)$ and $\rho_H^{an}/B = \rho_H^{an}/B(H, T_0, n)$ are the reduced amplitudes of contributions $\rho_{H0}$ and $\rho_H^{an}(\varphi)$ (see. Fig. 7), which depend on temperature and direction of the normal **n** to the sample surface; $R_H^0$ and $R_H^{an}$ are components of the ordinary Hall effect, connected with magnetic induction B, and $R_M^0$, $R_M^{an}$ are the coefficients of anomalous Hall effect determined by magnetization M and related to ferromagnetic component (see Figs. 3d-3f). Obviously, there are two independent ordinary contributions to Hall resistivity $R_H^0 \cdot B$ and $R_H^{an} \cdot B$, as well as two independent anomalous (ferromagnetic) components $R_M^0 \cdot 4\pi M$ and $R_M^{an} \cdot 4\pi M$, which differ for various **H** directions. Note that the last two components $R_H^{an} \cdot B$ and $R_M^{an} \cdot 4\pi M$ are responsible for the observed Hall effect anisotropy.

Fig. 11 shows the linear approximation within the framework of Eq. (6) of the reduced amplitudes $\rho_{H0}/B$ and $\rho_H^{an}/B$ *vs* $4\pi M/B$ in the range $T_0 = 2.1 - 6.5$ K for Ho$_{0.8}$Lu$_{0.2}$B$_{12}$ crystals with **n**∥[001], **n**∥[110] and **n**∥[111]. The linear fits are acceptable for directions **n**∥[001] and **n**∥[110] at temperatures of 2.1 K and 4.2 K for both the isotropic $\rho_{H0}/B$ and anisotropic $\rho_H^{an}/B$ contributions. The approximation was found to be valid at T = 6.5 K in the interval of small $4\pi M/B$ values (i.e., in high magnetic fields, in more detail see Fig. S4 in [52]), and the estimated parameters $R_H^0$ and $R_H^{an}$ are extracted as cutoffs, and $R_M^0$ and $R_M^{an}$ are the slopes of corresponding straight lines in Fig. 11. Since the parameters $R_H^0$, $R_H^{an}$, $R_M^0$, $R_M^{an}$ depend weakly on temperature (see Fig.S4 in [52]), the temperature-averaged values of these coefficients are summarized in Table 1.

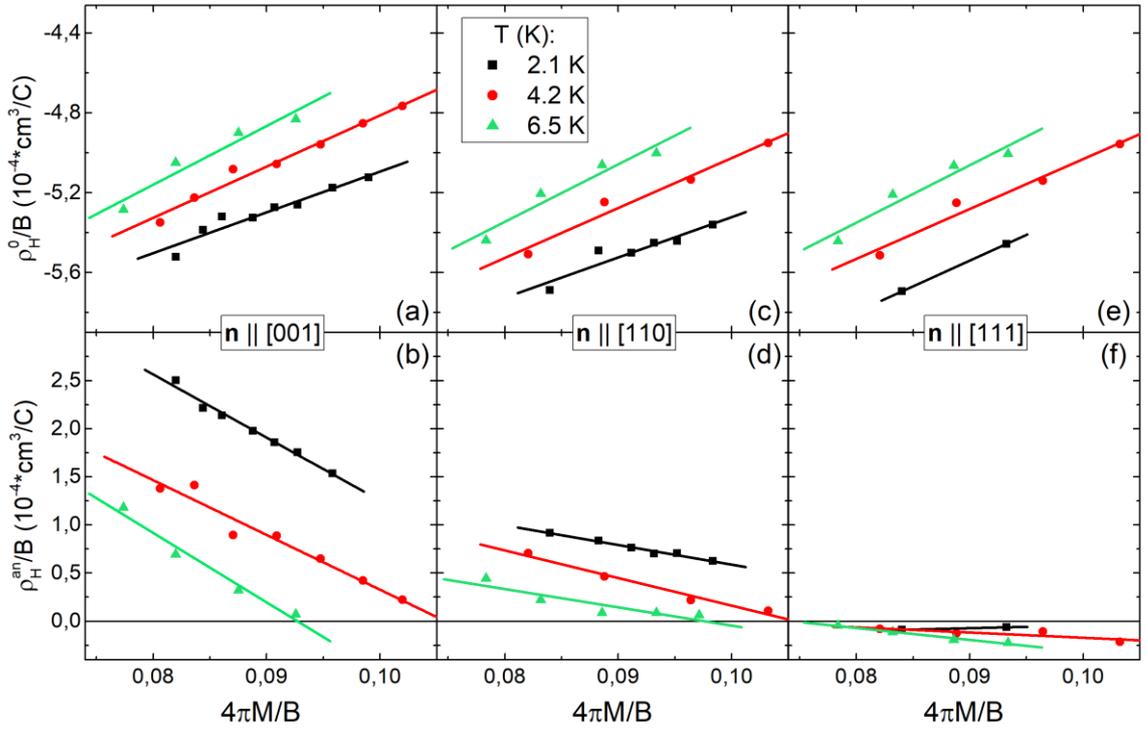

**Fig. 11.** Linear approximation of the isotropic $\rho_{H0}/B$ and anisotropic $\rho_H^{an}/B$ contributions *vs* $4\pi M/B$ within of Eq. (6) at 2-7 K for three principal directions **H**||[001], **H**||[110] and **H**||[111] (see text).

**Table 1.** Parameters $R_H^0$, $R_H^{an}$ of the ordinary and anomalous $R_M^0$, $R_M^{an}$ contributions to Hall effect in $Ho_{0.8}Lu_{0.2}B_{12}$ averaged over temperatures 2-7 K.

| $R_H$, $10^{-4}$*cm$^3$/C | **H** \|\| **n** \|\| [001] | **H** \|\| **n** \|\| [110] | **H** \|\| **n** \|\| [111] |
| --- | --- | --- | --- |
| $R_H^0$ | -7.3 | -7.7 | -7.7 |
| $R_H^{an}$ | 6.8 | 2.3 | 0 |
| $R_M^0$ | 25.2 | 26.9 | 26.5 |
| $R_M^{an}$ | -64.9 | -20.9 | 0 |

The analysis based on Eq.(6) allows to conclude that the values of coefficients $R_H^{an}$ and $R_M^{an}$, which are characteristics of the AHE turn out to be practically equal to zero for **H**||[111] (Fig. 11f). It can be seen that the Hall coefficient $R_H^0 \sim -7.5 \cdot 10^{-4}$ cm$^3$/C remains practically invariant for all **H** directions (see Table 1 and Fig. S4 in [52]), confirming that this ordinary negative component of Hall signal is about isotropic. On the contrary, the value of anisotropic positive contribution $R_H^{an}$ changes significantly from ~ $6.8 \cdot 10^{-4}$ cm$^3$/C for **H**‖[001] to $2.3 \cdot 10^{-4}$ cm$^3$/C in **H**‖[110] passing through zero for **H**‖[111] (Table 1). As a result, in Ho$_{0.8}$Lu$_{0.2}$B$_{12}$ for strong field **H**‖[001] this anisotropic positive component $R_H^{an}$ B·g(φ,n), which is proportional to magnetic induction, turns out to be comparable in absolute value with the negative isotropic component $R_{H0}$ B of ordinary Hall effect. Note that the values of coefficients $R_M^0$, $R_M^{an}$ of the anomalous (ferromagnetic) contributions, which are proportional to magnetization, exceed dramatically the ordinary parameters $R_H^0$, $R_H^{an}$ that agrees with the result [60] for Ho$_{0.5}$Lu$_{0.5}$B$_{12}$.

It is very unusual, that a significant isotropic anomalous positive contribution $R_M^0$~ $25 \div 27 \cdot 10^{-4}$ cm$^3$/C appears in the paramagnetic phase of Ho$_{0.8}$Lu$_{0.2}$B$_{12}$, and it may be attributed to the isotropic ferromagnetic component in the Hall signal. We propose that the $R_M^0$ term could be considered as a characteristics of the regime of ferromagnetic fluctuations detected above T$_N$ in the low field magnetic susceptibility (Fig. 4). On the contrary, the anomalous negative contribution $R_M^{an}$ varies strongly in the range (-) $21 \div 65 \cdot 10^{-4}$ cm$^3$/C depending from **M** direction (Table 1), and the component appears in strong magnetic field and at low temperatures (Fig.11). To summarize, AHE in the paramagnetic state of Ho$_{0.8}$Lu$_{0.2}$B$_{12}$ is proportional to magnetization and it is determined both by the positive contribution $R_M^0$ ·4πM·cos(φ) and by the strongly anisotropic component $R_M^{an}$·4πM·g(φ). These two contributions compensate each other in vicinity of **n**‖[110] (dynamic charge stripe direction [42,44]).

When discussing the nature of multicomponent Hall effect in Ho$_{0.8}$Lu$_{0.2}$B$_{12}$ it is worth noting the complicated multi-*q* incommensurate magnetic structure in the Neel state, which is characterized by propagation vector *q* = (1/2±δ, 1/2±δ, 1/2±δ) with δ = 0.035 and was detected in [49,63,64] for Ho$^{11}$B$_{12}$ in the neutron diffraction experiments at low temperatures in low (*H* < 20 kOe) magnetic field. It was found also for HoB$_{12}$ [49,63,64] that as the strength of external magnetic field increases above 20 kOe, the 4*q*-magnetic structure transforms into a more complex one, in which, apart from the coexistence of two AF 4*q* and *2q* components, there additionally arises some ferromagnetic order parameter. Then, a strong modulation of the diffuse neutron-scattering patterns was observed in HoB$_{12}$ well above $T_N$ [49,64] with broad peaks at positions of former magnetic reflections, e.g., at (3/2, 3/2, 3/2), pointing to strong correlations between the magnetic moments of Ho$^{3+}$ ions. These diffuse scattering patterns in the paramagnetic state have been explained in [49,64] by the appearance of correlated 1D spin chains (short chains of Ho$^{3+}$-ion moments placed on space diagonals <111> of the elementary unit), similar to those detected in low dimensional magnets [65]. It was found that these patterns can be resolved both well above (up to 70 K) and below $T_N$, where the 1D chains seem to condense into an ordered antiferromagnetic modulated (AFM) structure [49,64,66]. The authors [49] discussed the following scenario for the occurrence of long-range order in HoB$_{12}$: Far above $T_N$, strong interactions lead to correlations along [111], they are essentially one-dimensional and would not lead to long-range order at finite temperature. As $T_N$ is approached, the 1D-correlated regions grow in the perpendicular directions, possibly due to other interactions. *Cigar-shaped AFM-correlated regions* were proposed in [49] that become more spherical when $T_N$ is approached. Within this picture, the ordering temperature is located in the point where spherical symmetry is reached. Only then 3D behavior sets in, and HoB$_{12}$ exhibits long-range AFM order [49]. The refinement of Ho$^{11}$B$_{12}$ crystal structure was done with high accuracy in the space group Fm-3m, but also small static Jahn-Teller distortions were found in RB$_{12}$ compounds [31,36]. However, the most important factor of symmetry breaking is the dynamic one [31,36], which includes the formation both of vibrationally coupled Ho-Ho dimers and dynamic charge stripes (see [31,44,48] for more details). As a result, twofold symmetry in the (110) plane is conserved as expected for cubic crystal, but the charge stripes and Ho-Ho coupled vibrations suppress the exchange between nearest neighbored Ho-ions, resulting to emergence of complicated phase diagrams in the AF state with a number of different magnetic phases separated by radial and circular boundaries (Maltese Cross type of angular diagrams in RB$_{12}$ [37,39,41,42]). In this scenario AF magnetic fluctuations develop well above $T_N$ in HoB$_{12}$ along trigonal axis [111], and dynamic charge stripes along <110>

suppress dramatically the RKKY indirect exchange between Ho magnetic moments [39] provoking the formation of *cigar-shaped AFM-correlated regions* proposed in [49]. In our opinion, these effects are responsible both for the emergence of filamentary structures of fluctuating charges in these non-equilibrium metals and the formation of spin polarization in the conduction band, which results also to the appearance of a complicated multi-component Hall effect including two (isotropic and anisotropic) anomalous contributions.

### 4.2. Mechanisms of AHE in $Ho_{0.8}Lu_{0.2}B_{12}$.

Returning to commonly used classification [47], it is necessary to distinguish between the *intrinsic* AHE, related to the transverse velocity addition due to Berry phase contribution in systems with strong spin-orbit interaction (SOC), and the *extrinsic* AHE associated with scattering of charge carriers by impurity centers. However, AHE also arises in noncollinear ferromagnets, in which a nonzero scalar chirality $S_i(S_j \times S_k) \neq 0$ leads to the appearance of an effective magnetic field even in the absence of SOC [67], and in magnetic metals with a nontrivial topology of spin structures in real space [68–72]. When interpreting experimental data, the problem of identifying the actual mechanisms of AHE arises [47]. Skew scattering, for which the scattering angularly depends on the mutual orientation of the charge carrier spin and the magnetic moment of the impurity, predicts a linear relationship between the anomalous component of the resistivity tensor $\rho^{an}_H \sim \rho_{xx}$ and usually corresponds to the case of pure metals ($\rho_{xx} < 1$ μOhm·cm) [73]. In the range of resistances $\rho_{xx} = 1–100$ μOhm·cm the intrinsic AHE dominates, which is due to the effect of the Berry phase ($\rho^{an}_H \sim \rho_{xx}^2$) [74]. The contribution to scattering due to side jumping [75] with a similar scaling ($\rho^{an}_H \sim \rho^2_{xx0}$, where $\rho_{xx0}$ is the residual resistivity of the metal) is usually neglected [47]. In the "dirty limit" ($\rho_{xx} > 100$ μOhm*cm), an intermediate behavior is observed with a dependence of the form $\rho^{an}_H \sim \rho_{xx}^\beta$ and with an exponent $\beta = 1.6–1.8$, which is associated usually with the transition to hopping conductivity [47].

When identifying the AHE mechanism in $Ho_{0.8}Lu_{0.2}B_{12}$ with a small residual resistivity ($\rho_{xx0}$ ~1 µOhm·cm, Fig.1a), it is not possible to follow the traditional classification as no presence of the itinerant AHE with asymptotic $\rho^{an}_{xy} \sim \rho_{xx}^2$, or a skew scattering regime ($\rho^{an}_{H} \sim \rho_{xx}$) was found here. However, in order to correctly compare these diagonal and off-diagonal components of the resistivity tensor, it is possible to extract the corresponding anomalous contributions. As can be seen from Fig. 2, for **H**||**n**||[111] the anisotropic anomalous component of Hall signal is negligible and, as a result, the reduced Hall resistivity measured in angular experiments in direction **H**||**n** consists of the isotropic contribution only. In addition, according to conclusions made in [38,43], at low temperatures in paramagnetic state the magnetoresistance of $Ho_{0.8}Lu_{0.2}B_{12}$ consists of isotropic negative and anisotropic positive contributions, the latter being close to zero in the direction **H**||**n**||[111].

In this situation, for estimating the anisotropic components $\rho^{an}_{H}$ and $\rho^{an}_{xx}$, e.g., for **n**||[001] sample, it suffices to find the difference $\rho^{an}_{H}(\mathbf{n}||[001]) = \rho_{H}(\mathbf{n}||[001]) - \rho_{H}(\mathbf{n}||[111])$ (see Fig. 2b) and $\rho^{an}_{xx}(\mathbf{n}||[001]) = \rho(\mathbf{n}||[001]) - \rho(\mathbf{n}||[111])$ (see Fig. 2a). Fig. 12 demonstrates a scaling relation between these anisotropic components of $\rho^{an}_{H}$ and $\rho^{an}_{xx}$ for **H**||[001] and **H**||[110] directions, which leads to following conclusions. (*i*) For **H**||[001] a $\rho^{an}_{H} \sim \rho^{an\,1.7}_{xx}$ dependence is observed over the entire temperature range T ≤ T* ~ 60 K. This regime does not correspond to intrinsic AHE ($\beta$<2), and an onset of hopping conductivity ($\beta$ = 1.6–1.8) [47] seems to be an unreal scenario in this good metal ($\rho_{xx}$ ~1 µOhm·cm, Fig.1a). (*ii*) On the contrary, for **H**||[110], two anisotropic components of the resistivity tensor appear in the interval T < $T_S$ ~15 K and turn out to be related to each other as $\rho^{an}_{H} \sim \rho^{an\,0.83}_{xx}$ (Fig. 12), which does not favor skew scattering ($\beta$~1) [73]. Note that the exponent $\beta$ for **H**||[110] is twice smaller than that for **H**||[100] in $Ho_{0.8}Lu_{0.2}B_{12}$, and these regimes are observed in

adjacent $\rho^{an}_{xx}$ intervals changing one to another at $\rho^{an}_{xx} \sim 0.1$ µOhm·cm (Fig.12). Such a different behavior of charge transport parameters for two different magnetic field directions suggests that the AHE is caused by another scattering mechanism, which, in particular, may result from the influence of external magnetic field on dynamic charge stripes directed along <110> (see Fig. 1a). In this scenario the appearance of two types of AHE in $Ho_{0.8}Lu_{0.2}B_{12}$ may be interpreted as follows. The first mode of AHE associated with charge scattering in the interval $T < T_S \sim 15$ K is detected when magnetic field is applied along charge stripes (**H**||[110]), and the regime appears due to the formation of an infinite cluster in the filamentary structure of fluctuating charges. The second mode of AHE is induced by the order-disorder transition at T* ~ 60 K and corresponds to the magnetic field applied transverse to vibrationally coupled dimers of rare-earth ions (**H**||[001] ⊥ [110]). In the latter case, when the carrier moves in transverse magnetic field along a complex path, the intrinsic AHE is expected to be influenced by the Berry phase in real space [74], but instead, the $\rho^{an}_H \sim \rho^{an}_{xx}{}^{1.7}$ scaling is observed. This unusual behavior seems to be a challenge to the contemporary AHE theory and has to be clarified in future studies.

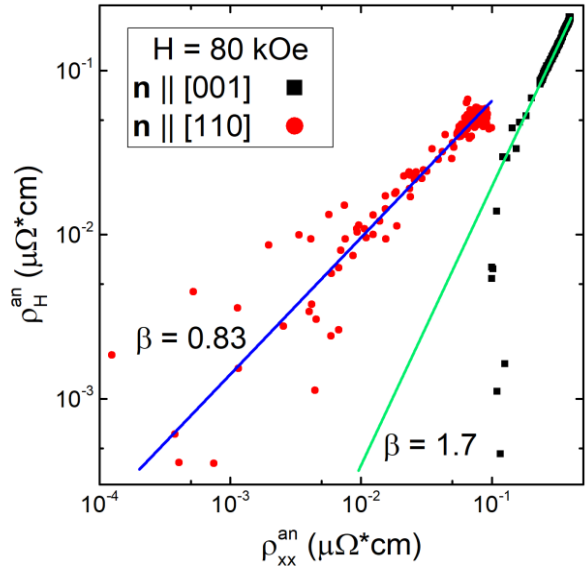

**Fig. 12.** Anisotropic AHE components for directions **H**||[001] and **H**||[110] in the magnetic field H = 80 kOe scaled in double logarithmic plot. Solid lines display the linear approximations and $\beta$ denotes the exponent in $\rho^{an}_H \sim \rho^{an}_{xx}{}^\beta$.

### 4.3. AHE anisotropy and dynamic charge stripes.

The above analysis of Hall effect contributions in $Ho_{0.8}Lu_{0.2}B_{12}$, based on measurements in the traditional scheme (Fig. 2, Fig. 3a-3c and Fig. S1a in [52]), and on the studies of angular dependences with applied magnetic field rotating in the (110) plane (Figs. 5-6 and Figs. S1b, S2 and S3 in [52]), allows to obtain a set of AHE coefficients, which characterize the ordinary and anomalous contributions *along three principal directions of the magnetic field* (**H**∥**n**∥[001], **H**∥**n**∥[110] and **H**∥**n**∥[111]).

In this case, the methodological feature of the performed angular measurements of Hall resistivity shows a cosine modulation of the projection of transverse Hall electric field on the direction that connects two Hall probes and is perpendicular to any of the specific normals (**n**∥[001], **n**∥[110] **n**∥[111], or **n**∥[211]). In this situation is more convenient to control the projection of the external field **H** onto the normal vector $H_n=(\mathbf{H}\cdot\mathbf{n})=H_0\cdot\cos\varphi$, which is used to determine the amplitude of contributions in the corresponding **n** direction (see inset in Fig. 2a). As can be seen from Fig. 10, vanishing of the AHE for **H** directed precisely along **n**∥[111] and **n**∥[112] does not mean zero values of $\rho_H^{an}(\varphi)/H$ for these crystals in the entire range of angles. In this case, it is obvious that for **H** in the plane of the sample, near zero values of $H_n=H_0\cdot\cos\varphi=0$, one should also expect zero values of anomalous contributions to Hall signal.

In [39,43–45,55] it was found that the angular dependence of magnetoresistance (MR) in $RB_{12}$ is determined by scattering of carriers on dynamic charge stripes and, as a result, the maximum positive values of MR are observed for **H**∥[001] perpendicular to the direction of these electron density fluctuations, while the minimum MR is observed for **H**∥[111] (see Fig. 13b). To clarify the nature of these anomalies in angular AHE curves, one can restore the angular dependence of the AHE in the entire range of 0-360° and compare the obtained curve with the related MR data, relying on experimental $\rho_H/B(\varphi)$ curves measured at T = 2.1 K in magnetic field H = 80 kOe for four different crystals when **H** is rotated in the same plane (110). Since both the ordinary and anomalous components of Hall signal can be described by cosine dependence $\rho_H(\varphi) = \rho_{H0}\cdot\cos(\varphi-\varphi_1)+\rho_H^{an}\cdot g(\varphi)$, the representation of experimental data shown in Fig. 13a in the form of $\rho_H(\varphi)/(H\cdot\cos(\varphi-\varphi_1))$ allows us to separate the isotropic and anisotropic contributions from Hall experiments.

The averaged envelope (indicated by yellow shading in Fig. 13a) was obtained after removing the particular portions of the related angular dependences with singularities associated with division by small values, and then averaging the data of these four angular Hall signal dependences. In this case, in accordance with the data in Fig. 2b, in **H**||<111> direction on the resulting envelope curve $\rho_H(\varphi)/(H\cdot\cos(\varphi-\varphi_1))$, the maximum negative values of about $-6\cdot10^{-4}$ cm$^3$/C correspond to isotropic ordinary component $\rho_{H0}/H$ of Hall effect, which is independent on magnetic field direction. The positive anisotropic component reconstructed from the data of four measurements (the yellow shading in Fig. 13a) provide changes of the $\rho_H(\varphi)/(H\cdot\cos(\varphi-\varphi_1))$ in the range $-(3.2\div6)\cdot10^{-4}$ cm$^3$/C. The location of its extrema coincide with the positions of anomalies on the MR curve (Fig. 13b). Indeed, the maximum positive contribution to AHE appears synchronously with the MR peak along <001>, while for <110>, a small (if compared with the anomaly along <001>) positive AHE component is recorded (Fig. 13a) simultaneously with a small amplitude singularity of MR. We also note that two spatial diagonals <111> on the anisotropic contribution $\rho_H^{an}g(\varphi)/(H\cdot\cos(\varphi-\varphi_1))$ seem to be equivalent and show no hysteretic features. The observed behavior of Hall effect in Ho$_{0.8}$Lu$_{0.2}$B$_{12}$ agrees very well with symmetry lowering of the *fcc* structure of Ho$_{0.8}$Lu$_{0.2}$B$_{12}$ due to static and dynamic Jahn-Teller distortions [31].

Finally, the comparison of the angular dependences of MR and Hall effect in Ho$_{0.8}$Lu$_{0.2}$B$_{12}$ (Fig. 13) shows that, along with the normal isotropic contributions to the diagonal (negative MR) and off-diagonal (the ordinary Hall coefficient of negative sign) components of the resistivity tensor, anomalous anisotropic positive components appear both in the MR and Hall effect at low temperatures and reach (*i*) maximal values in the direction of magnetic field transverse to dynamic charge stripes (**H**||[001]) and (*ii*) zero values for **H**||[111]. This anisotropy arises simultaneously with the transition to the cage glass state at T* ~ 60 K and seems to be related to the formation of vibrationally coupled pairs of rare-earth ions displaced from their centrosymmetric positions in the $B_{24}$ cavities [44]. The significant increase of this anisotropy is detected at temperatures $T < T_S \sim 15$ K upon the formation of a large size clusters in the filamentary structure of fluctuating electron density (stripes). Taking into account that, according to the results of room temperature measurements of the dynamic conductivity of LuB$_{12}$, about 70% of charge carriers participate in the formation of the collective mode (hot electrons) [76], a redistribution of carriers between the non-equilibrium and Drude components should be expected with decreasing temperature.

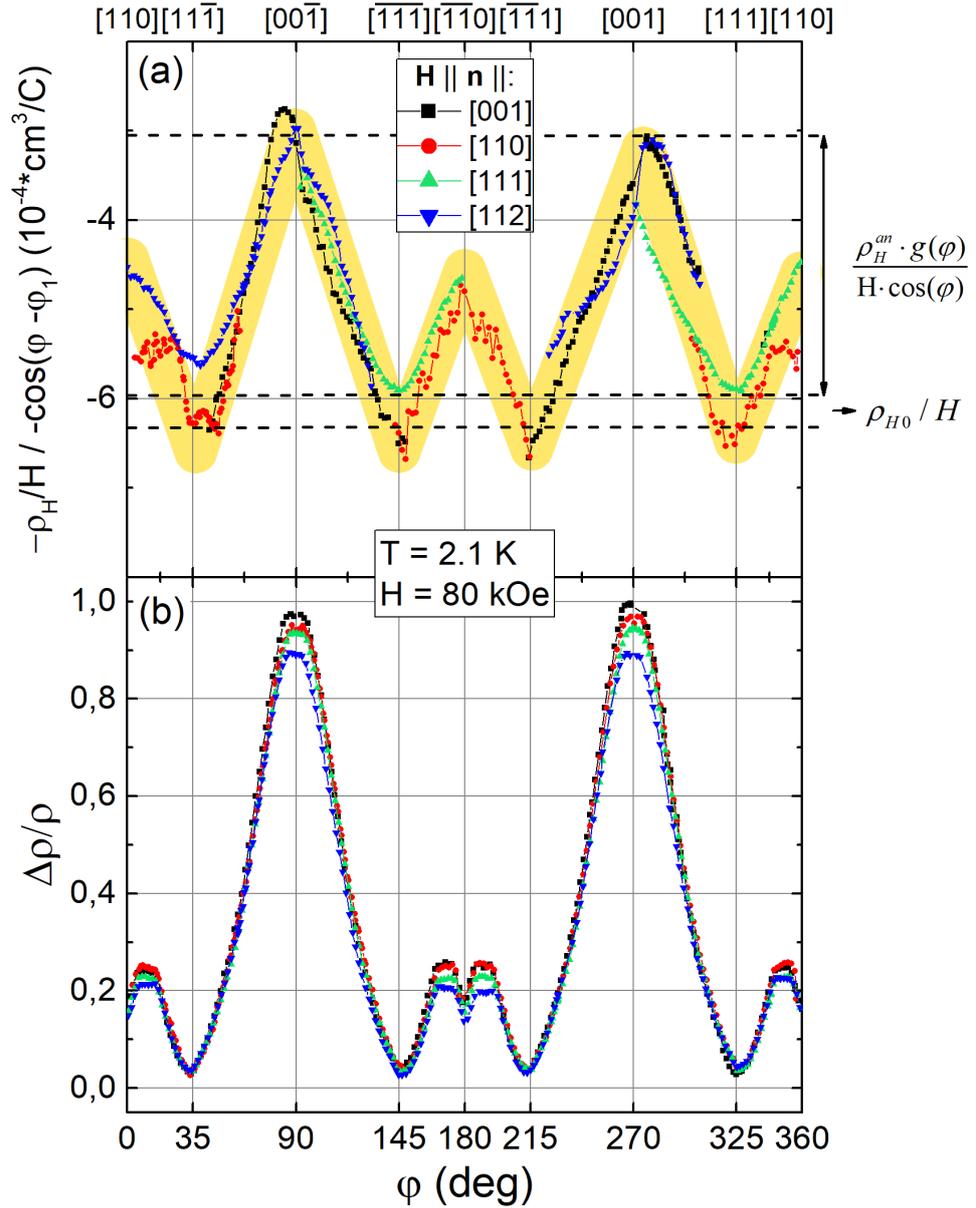

**Fig. 13.** (a) Normalized angular Hall resistivity $\rho_H(\varphi)/(H\cdot\cos(\varphi-\varphi_1))$ and (b) magnetoresistance curves, normalized to 1, in field of 80 kOe at temperature 2.1 K for four samples with **n**||[001], **n**||[110], **n**||[111] and **n**||[112]. Yellow shading indicates the common envelope for all four Hall effect measurements.

Apparently, the activation behavior of the Hall concentration of charge carriers observed in the range 60-300 K ($T_a \sim$ 14-17 K, Fig. 8a) may be attributed to the involvement of additional conduction electrons in the collective mode. When the vibrationally coupled dimers of rare-earth ions are formed below $T^* \sim 60$ K, chains of oriented stripes appear in magnetic field, initiating the emergence of intrinsic AHE (Fig. 12). We propose that AHE in Ho$_{0.8}$Lu$_{0.2}$B$_{12}$ is caused by the transverse addition to velocity due to the Berry phase contribution [47], which arises for carriers moving in a complex filamentary structure of the electron density in magnetic field applied transverse to dynamic stripes. During the formation of a large size clusters in the structure of stripes (interval $T < T_S \sim 15$K) in the field orientation along the stripes **H**||[110] no intrinsic AHE is expected in this configuration and the skew scattering contribution, for which the scattering angle depends on the mutual orientation of the charge carriers spin and the magnetic moment of the impurity, may become noticeable with a linear relationship $\rho^{an}_H \sim \rho^{an}_{xx}$ between these components of the resistivity tensor (Fig. 12). We propose that some geometric factors are responsible for the *β* exponent reduction in these two AHE regimes. Approaching the AF transition at $T \geq T_N$, on-site 4*f*-5*d* spin fluctuations in the vicinity of Ho$^{3+}$ ions lead to magnetic polarization of the 5*d* states of the conduction band, which gives rise to ferromagnetic fluctuations in Ho$_{0.8}$Lu$_{0.2}$B$_{12}$ (Fig. 4) producing ferromagnetic nano-size domains (ferrons), and, as a result, these initiate the appearance of the ferromagnetic contribution to AHE (Fig. 11 and Table 1). We emphasize that such a complex multicomponent AHE in the model Ho$_{0.8}$Lu$_{0.2}$B$_{12}$ compound with an *fcc* lattice turns out to be due to the inhomogeneity and complex filamentary structures of the electron density in the bulk of this model metal with the Jahn-Teller lattice instability and electron phase separation.

## 5. Conclusion.

In the paramagnetic phase of $Ho_{0.8}Lu_{0.2}B_{12}$ with cage-glass state below $T^* \sim 60$ K and electronic phase separation (dynamic charge stripes), magnetotransport properties were studied at temperatures 1.9–300 K in magnetic fields up to 80 kOe. Field and angular dependencies of resistivity and Hall resistivity measurements were performed on single-domain single-crystalline $Ho_{0.8}Lu_{0.2}B_{12}$ samples that allowed to separate and analyze several different contributions to the Hall effect. It was shown that, along with a negative ordinary isotropic component of Hall resistivity, an intrinsic AHE of a positive sign arises in the cage-glass state in the field direction $\mathbf{H}\|[001]$, which is perpendicular to the charge stripes. The AHE corresponds through to the relation $\rho^{an}_H \sim \rho^{an}_{xx}{}^{1.7}$ to the anomalous components of the resistivity tensor. It was also found that at temperatures $T < T_S \sim 15$ K, where a large size clusters develop in the filamentary structure of fluctuating charges (stripes), the contribution to AHE with a relationship of the form $\rho^{an}_H \sim \rho^{an}_{xx}{}^{0.83}$ becomes dominant when $\mathbf{H}\|[110]$. We propose that these two components are intrinsic (the transverse addition to the velocity due to the contribution of the Berry phase) and extrinsic (from the skew scattering mechanism) [47], respectively, with some geometric factor reduction of the exponents. In the paramagnetic phase near Neel temperature, on-site $4f$-$5d$ spin fluctuations in the vicinity of $Ho^{3+}$ ions were found to induce spin-polarized $5d$ states (ferromagnetic nano-size domains) in the conduction band of $Ho_{0.8}Lu_{0.2}B_{12}$, resulting to the appearance of additional ferromagnetic contribution to AHE, relating both the isotropic and anisotropic components of Hall effect. Detailed measurements of the angular dependencies of Hall resistivity and MR with the rotation of the vector $\mathbf{H}$ in the (110) plane, perpendicular to the direction of the stripes, made it possible to separate the negative isotropic and positive anisotropic contributions to AHE and MR, and explain them in terms of charge carriers scattering by dynamic charge stripes.

**Acknowledgements.** The work in Prokhorov General Physics Institute of RAS was supported by the Russian Science Foundation, Project No. 22-22-00243, and partly performed using the equipment of the Institute of Experimental Physics, Slovak Academy of Sciences. The work of K.F. and S.G. was supported by the Slovak Research and Development Agency under the contract No. APVV-17-0020, and by the projects VEGA 2/0032/20, and VA SR ITMS2014+313011W856.

of the angular dependencies of Hall resistance $\rho_H(\varphi)$ for H = 80 kOe in the temperature range 2.1 - 30 K for **n**||[111] and **n**||[112] by Eq. (2) (Figs.S2 and S3); coefficients $R_H^0$, $R_H^{an}$ of the ordinary and $R_M^0$, $R_M^{an}$ of the anomalous Hall effect (Eq.(6)) depending on temperature for three vectors **n** || [001], **n** || [110] and **n** || [111] (Fig.S4); corrections due to demagnetizing factors and values of the demagnetizing factor, depending on the type of experiment (Table S1).

# Supplementary information to the paper
## Hall effect anisotropy in the paramagnetic phase of $Ho_{0.8}Lu_{0.2}B_{12}$ induced by dynamic charge stripes

A.L. Khoroshilov[a*], K.M. Krasikov[a], A.N. Azarevich[a,b], A.V. Bogach[a], V.V. Glushkov[a], V.N. Krasnorussky[a,c], V.V. Voronov[a], N.Yu. Shitsevalova[d], V.B. Filipov[d], S. Gabáni[e], K. Flachbart[e], N. E. Sluchanko[a]

[a]*Prokhorov General Physics Institute of the Russian Academy of Sciences, Vavilova 38, 119991 Moscow, Russia*
[b]*Moscow Institute of Physics and Technology (State University), Moscow Region 141700 Russia*
[c]*Vereshchagin Institute for High Pressure Physics of RAS, 14 Kaluzhskoe Shosse, 142190 Troitsk, Russia*
[d]*Institute for Problems of Materials Science, NASU, Krzhizhanovsky str., 3, 03142 Kyiv, Ukraine*
[e]*Institute of Experimental Physics SAS, Watsonova 47, 04001 Košice, Slovakia*
*E-mail: artem.khoroshilov@phystech.edu*

**1. Relationship between the traditional technique for Hall effect measurement and the scheme with step-by-step sample rotation in external magnetic field.**

Hall resistivity of the samples was detected in two ways (see Figs.S1a and S1b below, correspondingly):

*(a)* The traditional technique for the Hall effect measurements was applied, where we calculate the value of Hall resistivity as $\rho_H/H = [(V_H(+H) – V_H(–H))/(2I)]\cdot d/H$ (I is the measuring current through the sample, d is the thickness of the sample, and $V_H(+/–H)$ are the voltages measured from Hall probes in two opposite directions of the external magnetic field $H \perp I$, and

*(b)* The angular dependences of Hall resistivity were obtained using a measuring cell of an original design, which provides the rotation of vector **H** located in the plane perpendicular to the current direction **I** || [1-10] with a minimum step $\Delta\varphi = 0.4°$ (see the schematic view on the inset in fig. 2a in the manuscript). Measurements were carried out in a wide temperature range 1.9 - 300 K in magnetic fields up to 80 kOe, the angle $\varphi = \mathbf{n}^\wedge\mathbf{H}$ between the direction of the normal **n** to the lateral surface of sample and external magnetic field **H** varied in the range $\varphi$ = 0 - 360°. The setup for measurements was equipped with a stepper motor with automatic control of stepwise rotation of sample. High accuracy of stabilization of temperature ($\Delta T \approx 0.002$ K in the range 1.9 - 7K) and magnetic field ($\Delta H \approx 2$ Oe) was ensured, respectively, by Cryotel TC 1.5/300 temperature controller and Cryotel SMPS 100 superconducting magnet power supply in combination with CERNOX 1050 thermometer and n-InSb Hall sensors.

Thus, the panel (a) in Fig. S1 below shows the magnetic field curves of the reduced Hall resistivity $-\rho_H/H(H)$ at T=2.1 K for H≤ 80 kOe. The panel (b) in Fig. S1 performs the angular dependencies of the reduced Hall resistivity $\rho_H/H(\varphi)$ in magnetic field H=80 kOe at T=2.1 K. The dotted lines indicate the correspondence between the data at H=80 kOe on the $-\rho_H/H(H)$ curves ($\varphi = 0°$; 180°) and the angular $\rho_H/H(\varphi)$ curves ($\varphi = 0° - 360°$). For convenience, in Fig. S1b, the curves for all samples with different normal vectors except **n**||[110] are shifted by corresponding angles, so that the maxima of the Hall signal are along the corresponding normal directions. The positions of these maxima are marked with arrows in Fig. S1b, the upper axis indices indicate the crystallographic directions of the corresponding **n** vectors.

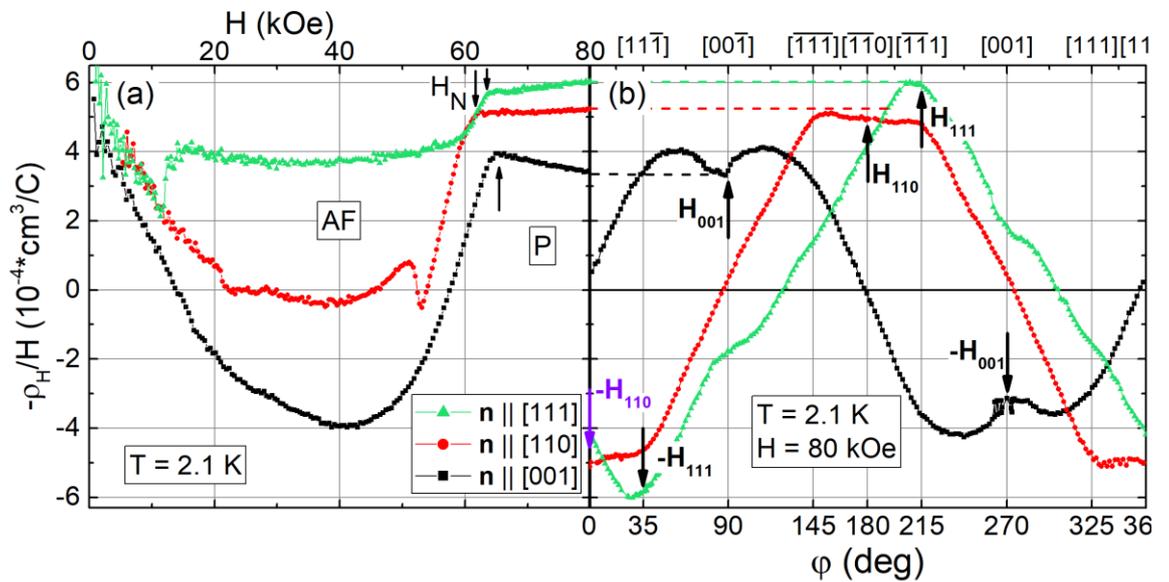

**Fig. S1.** Relationship between (a) the traditional scheme of Hall effect measurements with two opposite orientations of magnetic field strength ±**H**||**n** and (b) the scheme with step-by-step sample rotation in a plane perpendicular to the direction of the measuring current **I** || [1-10] (see also inset in fig. 2a of the paper).

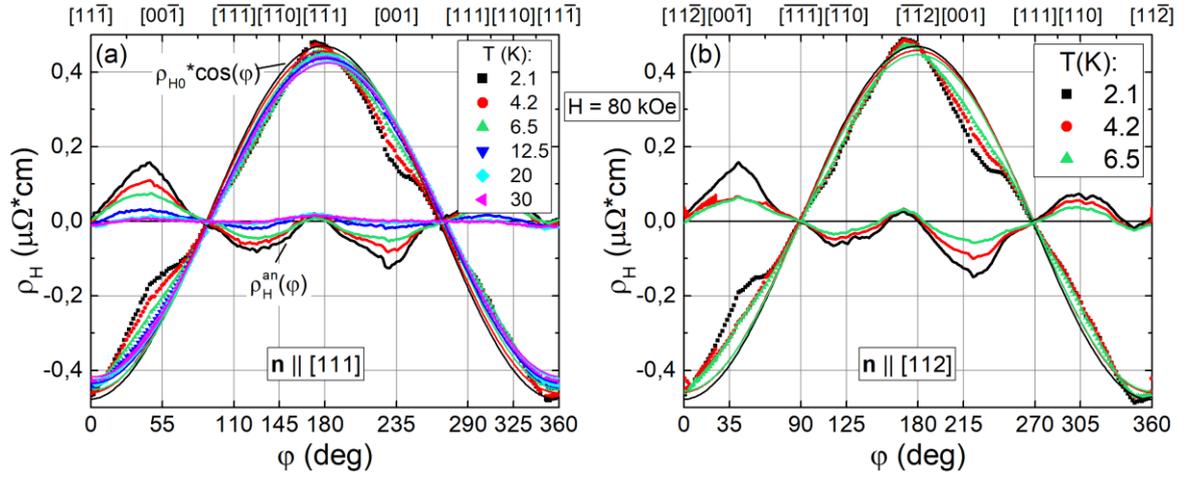

**Fig. S2.** (a) - (b) Approximation by Eq.(2) in the article of the angular dependencies of Hall resistivity $\rho_H(\varphi)$ for H = 80 kOe in the temperature range 2.1 - 30 K for **n**‖[111] and **n**‖[112]. Symbols show the experimental values of $\rho_H(\varphi)$, thin curves present the harmonic contribution $f_{cos}(\varphi) \approx \rho_{H0} \cdot \cos(\varphi)$, thick curves demonstrate the anharmonic component $\rho_H^{an}(\varphi) = \rho_H(\varphi) - f_{cos}(\varphi)$.

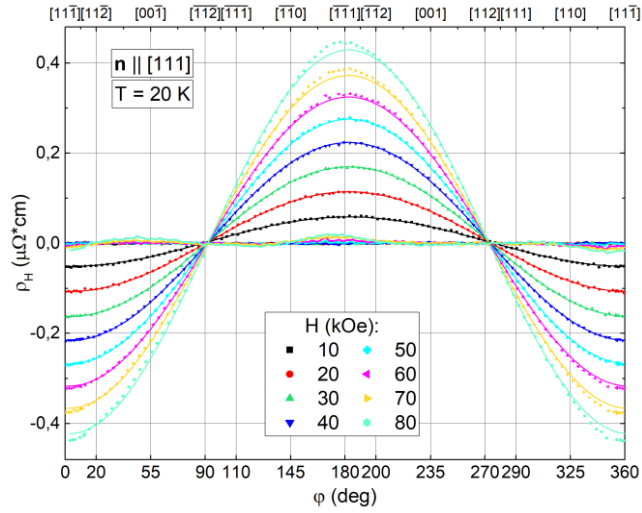

**Fig. S3.** Approximation of the angular dependences of Hall resistivity $\rho_H(\varphi)$ at T = 20 K in magnetic fields up to 80 kOe for **n**‖[111]. Symbols show the experimental values of $\rho_H(\varphi)$, thin curves present the harmonic contribution $f_{cos}(\varphi) \approx \rho_{H0} \cdot \cos(\varphi)$, thick curves demonstrate the anharmonic component $\rho_H^{an}(\varphi) = \rho_H(\varphi) - f_{cos}(\varphi)$.

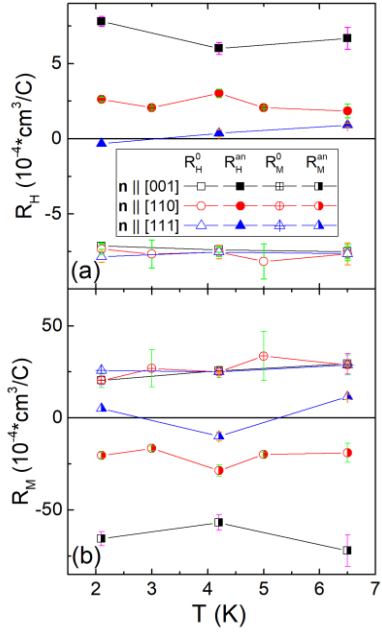

**Fig. S4.** Coefficients (a) $R_H^0$, $R_H^{an}$ of the ordinary and (b) $R_M^0$, $R_M^{an}$ of the anomalous Hall effect depending on temperature for three normal vectors **n** ∥ [001], **n** ∥ [110] and **n** ∥ [111].

**2. Corrections of the M/B(B) data due to demagnetizing factors, to match the experimental data of $\rho_H$/B(B) and M/B(B).**

Measurements of the field dependencies of magnetization **M**(H) and the Hall resistivity $\rho_H$(H) were carried out on samples cut from the same single crystal, but having different geometric shapes. Moreover, the direction of external magnetic field always coincided with the direction of the sample lateral surface normal vector **H**∥**n**. Therefore, for more accurate calculation of the magnetic field induction **B** in the sample, we used a two-stage calculation with two different demagnetizing factors. They corresponded to the magnetization experiment (N1) and to the Hall resistivity experiment (N2), respectively (see Table S1 below).

**Table S1.** Values of the demagnetizing factor N, depending on the type of experiment performed and the direction of the normal vector **n** to the sample surface.

|  | **n** ∥ [001] | **n** ∥ [110] | **n** ∥ [111] |
|---|---|---|---|
| $N_1$, M(H) experiments | 0.3972 | 0.3882 | 0.1734 |
| $N_2$, $\rho_H$(H) experiments | 0.1765 | 0.5324 | 0.54095 |

The maximum value of the magnetic field strength **H** in the Hall resistivity $\rho_H(H)$ measurements was 80 kOe, while when measuring the magnitude of **M**(H) this value reached 70 kOe in experiments with samples **n**||[001] and **n**||[110], and 50 kOe for the experiment with sample **n**||[111]. In order to match the obtained arrays of curves, the available data of magnetization **M**(H) were linearly approximated in the range 70 - 80 kOe for all three directions of **n**.